\documentclass[12pt,a4paper]{article}

\usepackage[greek,american]{babel}
\usepackage{amssymb}
\usepackage{amsfonts}
\usepackage{amsmath}
\usepackage{relsize}
\usepackage{subfigure}
\usepackage[pdftex]{graphicx}

\graphicspath{{./art/}}
\usepackage{float}
\usepackage{adjustbox}

\newcommand{\iid}{\stackrel{{\rm i.i.d.}}\sim}
\newcommand{\ind}{\stackrel{{\rm ind.}}\sim}

\newcommand{\given}{\,|\,}

\def\G{\mathbb G}
\def\R{\mathbb R}

\def\Q{\mathbb Q}
\def\E{\mathbb E}
\def\X{\mathbb X}

\def\X{\mathbb X}

\def\bl{\boldsymbol}

\def\l{\lambda}
\def\a{\alpha}
\def\b{\beta}
\def\s{\sigma}

\def\d{\delta}
\def\t{\theta}

\renewcommand{\vec}[1]{\mathbf{#1}}

\usepackage{natbib}
\usepackage[autostyle=true]{csquotes} 

\begin{document}

\begin{center}

{{{\Large\sf\bf Joint reconstruction and prediction of random dynamical systems under borrowing of strength}}}\\

\vspace{0.5cm}
{\large\sf Spyridon J. Hatjispyros
     \footnote{ Corresponding author. Tel.:+30 22730 82.326\\
     \indent E-mail address: schatz@aegean.gr },
      Christos Merkatas}

\vspace{0.2cm}
\end{center}
\centerline{ \sf Department of Mathematics, Division of Statistics and Actuarial Science, University of the Aegean}
\centerline{\sf  Karlovassi, Samos, GR-832 00, Greece.} 

\begin{abstract}
We propose a Bayesian nonparametric model based on Markov Chain Monte Carlo (MCMC) methods 
for the joint reconstruction and prediction of discrete  time stochastic dynamical systems, based on $m$-multiple time-series data, perturbed by additive dynamical noise.
We introduce the Pairwise Dependent Geometric Stick-Breaking Reconstruction (PD-GSBR)
model, which relies on the construction of a $m$-variate nonparametric prior over the space of 
densities supported over $\R^m$. 
We are focusing in the case where at least
one of the time-series has a sufficiently large sample size representation
for an independent and accurate Geometric 
Stick-Breaking estimation, as defined in Merkatas et al. (2017). 
Our contention, is that whenever the dynamical error processes perturbing the 
underlying dynamical systems share common characteristics, 
underrepresented data sets can benefit in terms of model estimation accuracy. 
The PD-GSBR estimation and prediction procedure is demonstrated 
specifically in the case of maps with polynomial nonlinearities of an arbitrary degree. 
Simulations based on synthetic time-series are presented.

\vspace{0.1in} 
\end{abstract}

\section{Introduction}

The interdisciplinary framework of nonlinear dynamical systems, has been used extensively 
for the modeling of time varying phenomena in physics, chemistry, biology, economics  
and so forth, that they exhibit complex and irregular behavior \citep{ott2002chaos}.  
The apparently random and unpredictable behavior of deterministic chaotic dynamics, 
right from the early days of the theory, prompted to the use of random or probabilistic methods \citep{berliner1992statistics, chatterjee1992chaos}. At the same time, the ubiquitous effect of 
the different kinds of noise in experimental or real data, reinforced the interaction between 
nonlinear dynamics and statistics \citep{mees2012nonlinear}. 

When the nonlinear procedure is influenced  by the uncertainty of the measurement process, 
the resulting time-series can be thought of as the corruption of the true system states by 
observational noise. We can consider the noise in this case, as being added after the 
time evolution of the trajectories under consideration, thus inducing a blurring effect on the 
true evolution of the process. In this case the dynamics of the process are not influenced, 
and the invariant measure of the process is the convolution of the unperturbed measure and 
the noise distribution. Observational noise corrupted dynamical systems are often confronted 
with time delay embedding techniques and related methods \citep{Ruelle,abarbanel2012analysis,kantz2004nonlinear}.

In the case of dynamical or interactive noise, the noise is incorporated at each step 
of time evolution of the trajectories. 
For example consider a situation in which at each discrete time, the state 
of the system is reached with some error. Then the constructed predictive model consists of two
parts, the nonlinear-deterministic component and the random noise. In such cases where the noise 
acts as a driving force, the underlying deterministic dynamics can be drastically modified \citep{jaeger1997homoclinic}, and the predictive model constitutes what is known as a
random dynamical system \citep{Arnold,smith2000nonlinear}. From a modeling perspective,
the existence of a stochastic forcing term can be thought of as representing the error in the 
assumed model, mimicking the aggregate action of variables not included in the model, 
compensating for a small number of degrees of freedom. In fact, when a small number of degrees 
of freedom is segregated from a larger coupled system, usually has as an outcome, reduced equations 
with deterministic and stochastic components \citep{He2014ChaosNoise}. 

Methods based on deterministic inference in the case of dynamical noise are  inefficient,
and many methods have been proposed by various researchers to address the various aspects of the 
problem. A theorem formulated to cope with the embedding problem for random dynamical systems 
is given in \cite{muldoon1998delay}. In \cite{siegert1998analysis} and \cite{siefert2004reconstruction}
the issue of dynamical reconstruction is addressed for stochastic systems described by Langevin type
of equations under different types of random noise. Because of the different impact of the noise 
types, the goal of estimating the noise density directly from the data is highly significant \citep{heald2000estimation,strumik2008influence,siefert2004differentiate}.

The Bayesian framework \citep{robert2007bayesian} was initially put into context by Davies \citep{davies1998nonlinear} for nonlinear noise reduction. 
\cite{Meyer2000, Meyer2001} applied MCMC methods for the parametric estimation of state-space nonlinear models, extending maximum likelihood-based methods \citep{McSharry}. 
Later, \cite{smelyanskiy2005reconstruction} 
reconstructed stochastic nonlinear dynamical models from trajectory measurements, using
path integral representations for the likelihood function,
extended for nonstationary systems in \cite{luchinsky2008inferential}. 
\cite{matsumoto2001reconstructions} introduced a hierarchical Bayesian approach
and later, \cite{nakada2005bayesian} applied a hybrid Monte Carlo scheme
for the reconstruction and prediction of nonlinear dynamical systems.
More recently in \cite{molkov2012random}, a Bayesian technique was proposed for predicting the 
qualitative behavior of random dynamical systems from time-series.

In the literature of stochastically perturbed dynamical systems,
error processes are frequently modeled via zero mean Gaussian distributions.
Such an assumption, when violated, can cause inferential 
problems. For example when the noise process produces outlying errors the 
estimated variance under the normality assumption, is artificially enlarged causing 
poor inference for the system parameters. 
Alternatively, we could make the assumption of the existence of two sources of 
random perturbations. For example we could assume that an environmental source,
caused perhaps by spatiotemporal inhomogeneities \citep{strumik2008influence}, 
is producing weak and frequent perturbations. At the same time, stronger but less frequent 
perturbations in the form of outlying errors, are coming from a higher dimensional 
deterministic component.
Other cases could include systems exerting noise at random time intervals and impulsive noise  \citep{shinde1974signal,middleton1977statistical}. These are situations where the noise probability 
density function does not decay in the tails like a Gaussian. Also, when the system 
under consideration is coupled to multiple stochastic environments 
the driving noise term may exhibit non-Gaussian behavior, see for example references \cite{kanazawa2015minimal} and \cite{kanazawa2015asymptotic}.

A number of approaches for modeling time-series in a Bayesian {\it nonparametric} context have been proposed in the literature. For example, an infinite mixture of time-series models has been proposed in \citet{rodriguez2008bayesian}. A Markov-switching finite mixture of independent Dirichlet process
mixtures has been proposed by \citet{taddy2009markov}. More recently, \citet{jensen2010bayesian} and \citet{griffin2010inference}, considered 
Dirichlet process mixtures for stochastic volatility models in discrete and continuous time, respectively. An approach for continuous time-series modeling based on time dependent Geometric Stick-Breaking process
mixtures can be found in \cite{mena2011geometric}. 
For a Bayesian nonparametric nonlinear noise reduction approach see 
\cite{kaloudis2018denoising}.

Recently there has been a growing research interest for Bayesian nonparametric modeling in the context of {\it multiple} time-series. In \cite{fox2009sharing} a Bayesian nonparametric model based on the Beta process was introduced in order to model dynamical behavior shared among a number of time-series. They represented the behavioral set with an attribute list encoded by an $n\times k$ binary matrix, with $n$ the number of time-series and $k$ the number of features. Their approach allowed for potentially an infinite number of behaviors $k$. This was an improvement of a similar approach of a previous work \cite{fox2008hdp} where the time-series shared exactly the same set of behaviors.
In \cite{nieto2016bayesian}, a Bayesian nonparametric dynamic autoregressive model for the analysis of multiple time-series was introduced. They considered an autoregressive model of order $p$ for each of the time-series in the collection, and a Bayesian nonparametric prior based on dependent P\'olya trees. The dependent prior, with its median fixed at zero, was used for the modeling of the errors. 

In a previous work \citet{merkatas2017bayesian} we have dealt with the problem of identifying the
deterministic part of a stochastic dynamical system and the 
estimation of the associated unknown density of dynamical perturbations, which is perhaps
non-Gaussian, simultaneously from data via Geometric Stick-Breaking (GSB) mixture processes. 
In this work we will attempt to generalize the so-called Geometric Stick-Breaking Reconstruction (GSBR) model under a multidimensional setting, in order to reconstruct and predict jointly, an arbitrary number of discrete time dynamical systems. More specifically, given a collection of $m$ noisy chaotic 
time-series, we propose a Bayesian nonparametric mixture model for the joint 
reconstruction and prediction of $m$ dynamical equations. 

Our method of joint reconstruction and prediction, is primarily based on the existence of a 
multivariate Bayesian nonparametric prior over the collection of the unknown dynamical noise 
processes. It is based on 
the Pairwise Dependent Geometric Stick-Breaking Process mixture priors developed in
\cite{hatjispyros2017dependent}, under the following assumptions:
\begin{enumerate}

\item
The dynamical equations have deterministic parts that they belong to known families of functions; for example they can be polynomial or rational functions.

\item
A-priori we assume that we have the knowledge that the noise processes corrupting dynamically 
the observed multiple time-series, have possibly common characteristics;
for example the error processes could reveal a similar tail behavior or (and) have common variances, or simply they come from the same noise process which is (perhaps) non-Gaussian.

\end{enumerate}
Our contention is that whenever there is at least one sufficiently large data set, using  
borrowing of strength prior specifications,  we will be able to recover the dynamical 
process for which we have insufficient information i.e. the process for which the sample size is inadequate for an independent GSBR reconstruction and prediction.

\smallskip
This paper is organized as follows. 
In Sec. II, we are giving some preliminary notions on the GSB mixture priors 
applied on a single discretized random dynamical system. 
In Sec. III, we introduce the joint probability model for the multiple time-series 
observations and we derive the PD-GSBR model.
We describe the associated joint nonparametric likelihood for the model, 
and as a special case we derive the joint parametric likelihood corresponding
to the assumption of common Gaussian noise along the multiple time-series 
observations. We also provide the PD-GSBR based Gibbs sampler for the estimation
of the unknown error processes, the control parameters, the initial conditions, 
and the out-of-sample predictions.
In Sec. IV, we resort to simulation. We apply the PD-GSBR model on the reconstruction 
and prediction of two pairs and one triple of random polynomial maps 
that are dynamically perturbed additively, by noise processes which are non-Gaussian. 
Finally, conclusions and directions for future research are discussed.

\section{Preliminaries}

For $j=1,\ldots,m$, we consider  the following assemblage of the $m$ decoupled random recurrences
\begin{align}
\label{dynamical_equations}
X_{ji} & = T_j(\t_j, X_{j,i-1},\ldots,X_{j,i-\rho_j},Z_{ji})\\
       & = g_j(\t_j, X_{j,i-1},\ldots,X_{j,i-\rho_j})+Z_{ji},\,\,i\ge 1,\nonumber
\end{align}
where $g_j:\Theta_j \times{\mathbb X}_j^{\rho_j}\to {\mathbb X}_j$, for some compact subsets 
${\mathbb X}_j$ of $\R$, $(X_{ji})_{i\,\ge\,1-\rho_j}$ and $(Z_{ji})_{i\,\ge\, 1}$ are real random
variables over some probability space $(\Omega,{\cal F},{\rm P})$; we denote by $\t_j\in\Theta_j
\subseteq\R^{q_j}$ any dependence of the deterministic map $g_j$ on parameters. 
$g_j$ is a nonlinear map, for simplicity continuous in the variable 
${\bf X}_{j, i:\rho_j}:=(X_{j,i-1},\ldots,X_{j,i-\rho_j})$.
We assume that the random variables $Z_{ji}$ are independent to each other,
and independent of the states $X_{j,i-r_j}$ for all $r_j < i + \rho_j$.
In addition, we assume that the additive perturbations $Z_{ji}$ are identically 
distributed from zero mean symmetric distributions with unknown densities $f_j$ defined 
over the real line so that $T_j:\Theta_j \times{\mathbb X}_j^{\rho_j}\times\R\to\R$.
Finally, notice that the lag-one stochastic process $(W_{ji}^1,\ldots,W_{ji}^{\rho_j})$
formed out by the time-delayed values of the $(X_{ji})$ process is Markovian over $\R^{\rho_j}$.

We assume that there is no observational noise. We denote the set of 
observations along the $m$ time-series as 
${\bl x}=\{ x_{ji}:1\le j\le m,1\le i\le n_j\}$ and with 
${\bl x}_j=\{ x_{ji}:1\le i\le n_j\}$ the set of observations in the 
$j$-th time-series. These are realizations of the nonlinear stochastic 
processes defined in (\ref{dynamical_equations}) for some unknown initial conditions 
${\boldsymbol x}_0=\{{\bl x}_{j,1:\rho_j}:1\le j\le m\}$.
The collection of the time-series observations $\boldsymbol x$, depends solely on the
initial distribution of the variable ${\boldsymbol X}_0$, the values of the control
parameters ${\boldsymbol \t}=\{\t_j:1\le j\le m\}$, and the particular realization of 
the noise processes.

In \cite{merkatas2017bayesian} a Bayesian nonparametric methodology is proposed
for the estimation and prediction of a single discretized random dynamical system 
from an observed noisy time-series of length $n$. It relaxes the assumption of normal perturbations
by assuming that the prior over the unknown density $f$ of the additive dynamical errors, 
is a random infinite mixture of zero mean Gaussian kernels. More specifically
a-priori we set
$$
f(z)=\int_{v>0}{\cal N}(z|0,v^{-1})\,\G(dv)=\sum_{k=1}^\infty\pi_k\,{\cal N}(z|0,\tau_k^{-1}),
$$
where $\G$ is a GSB random measure.
The random measure $\G$ is closely related to the well known Dirichlet random measure 
$\G'=\sum_{k\ge 1}w_k\d_{\tau_k}\sim{\cal DP}(c,G_0)$\citep{Ferguson,Sethuraman94}. $\d_{\tau_k}$'s 
are Dirac measures concentrated on the random precisions $\tau_k$'s, which in turn are
independently drawn (i.i.d.) from the mean parametric distribution $G_0$, being the prior guess
of $\G'$ i.e. $\E(G'(A))=G_0$ for measurable subsets $A$ of $\R_+$.
The probability-weights $w_k$ 
are stick-breaking in the sense that $w_1=v_1$ and $w_k=v_k\prod_{l<k}(1-v_l)$ and random 
because $v_k\iid{\cal B}e(1,c)$ a beta density with mean $(1+c)^{-1}$. 
We define the GSB random measure as 
$\G=\sum_{k\ge 1}\pi_k\d_{\tau_k}\sim{\cal GSB}(\a,\b,G_0)$ with $\pi_k=\E(w_k)$,
hence removing a hierarchy from the random measure $\G'$\citep{Fuentes}.
Then for $\l=(1+c)^{-1}$, we have $\pi_k=\l\,(1-\l)^{k-1}$ and $\E(G(A))=G_0$. 
Finally we randomize the probability-weights by letting $\l\sim{\cal B}e(\a,\b)$;
then $\l$ a-posteriori is again beta with its parameters 
updated by a sufficient statistic of the data. In \cite{merkatas2017bayesian}
it is shown that a $\G$-based Bayesian nonparametric framework for dynamical system estimation
is efficient, faster and less complicated
when compared to Bayesian nonparametric modeling via the Dirichlet process.

\smallskip
To sample from the posterior of $f(z)$, the control parameters of the 
deterministic part, the initial condition and the future observations, given the 
noisy time-series, in a finite number of steps we have to:
\begin{enumerate}

\item
Introduce the infinite mixture allocation variables $\{d_i:1\le i\le n\}$, 
such that P$\{d_i=k\}=\pi_k$ for $k\ge 1$, indicating the 
component of the infinite mixture the $i$th observation came from. 

\item
Augment the random density $f(z)$, with the auxiliary variables $\{N_i:1\le i\le n\}$, such
that the $N_i$'s are identically distributed from the specific negative binomial distribution $f_N(k|\l)={\cal NB}(k|2,\l)=k\l^2(1-\l)^{k-1}{\cal I}(k\ge 1)$. Then $d_i$ conditionally on 
$N_i$, attains the discrete uniform distribution over the random set $\{1,\ldots,N_i\}$. 

\end{enumerate}
Thereby, the dimension of the Gibbs sampler will be of order $\max_{1\le i\le n}N_i<\infty$.

\section{The PD-GSBR model}

We will model a-priori the errors $Z_j$ in the multiple recurrence relation
(\ref{dynamical_equations}) with a multivariate distribution over the space of densities. 
More specifically, we are interested in constructing for any finite integer $m\ge 2$
\begin{equation}
\label{errorvec1}
{\bl f}=(f_1,\ldots,f_m)^T,
\end{equation}
where $T$ denotes transposition, each $f_j$ is a random density function, and we 
are able to understand the dependence 
mechanism between pairs $(f_j, f_l)$ for each $j\neq l$. 

We will allow {\it pairwise} dependence between any two $f_j$ and $f_l$, 
so that there is a unique common component for each pair $(f_j, f_l)$. 
For example consider such a dependence structure
for $m=3$ for the random variables 
$Y_1=M_{11}\!+\!M_{12}\!+\!M_{13}$, $Y_2=M_{12}\!+\!M_{22}\!+\!M_{23}$ and
$Y_3=M_{13}+M_{23}+M_{33}$,
where all the $M_{jl}$ random variables are mutually independent. Then the dependence between $Y_j$
and $Y_l$ is created via them having $M_{jl}$ in common and it is easy to show that 
Cov$(Y_j,Y_l)=$Var$(M_{jl})$. Therefore, the independent
variables $M_{12},M_{13}$ and $M_{23}$ play the r\^ole of common parts for the pairs
$(Y_1,Y_2),(Y_1,Y_3)$ and $(Y_2,Y_3)$, respectively. On the other hand the 
independent variables $M_{11}, M_{22}$ and $M_{33}$ serve as idiosyncratic parts of the
variables $Y_1,Y_2$ and $Y_3$, respectively. In a more compact notation, we set 
${\bl Y}={\bl M}\!\cdot\!{\bl 1}$ where ${\bl Y}$ is the column vector of $Y_j$'s, ${\bl M}$
is a random {\it symmetric} matrix of independent random variables, and $\bl 1$ a $3\times1$ 
matrix of ones.
 
We will use this basic plan but instead of the real valued random vector $\bl Y$, 
we have the vector of random density functions $\bl f$. We set
${\bl f}=({\bl p}\otimes{\bl M})\cdot{\bl 1}$. In this case $\bl M$ is a $m\times m$
symmetric matrix of independent random zero mean mixture densities, 
$\bl p$ is a random stochastic matrix (its row elements add up to $1$ a.s.), 
and $\bl 1$ a $m\times1$ matrix of ones. The Hadamard product of the two matrices 
$\bl p$ and $\bl M$ is defined as $({\bl p}\otimes{\bl M})_{jl}=p_{jl}M_{jl}$, whence
$f_j=\sum_{l=1}^mp_{jl}\,M_{jl}$ with $M_{jl}=M_{lj}$
and ${\rm Cov}(f_j,f_l\,|\,{\bl p})=p_{jl}\,p_{lj}{\rm Var}(M_{jl})$.
We will model the densities
$f_j$ via
\begin{align}
& f_j(z)=\int\limits_{v>0}{\cal N}(z\,|\,0,v^{-1})\,\Q_j(dv),\,\,
  \Q_j=\sum_{l=1}^mp_{jl}\G_{jl}\nonumber\\
& \G_{jl}=\sum_{k=1}^\infty\pi_{jlk}\delta_{\tau_{jlk}}\ind{\cal GSB}(\a_{jl},\b_{jl},G_0),\,\,
   \G_{jl}=\G_{lj}\,\,{\rm a.s.},\nonumber
\end{align}
where for the random selection-probabilities $p_{jl}$ it is that $\sum_{l=1}^mp_{jl}=1$ a.s., 
and $\tau_{jlk}\iid G_0$. 
The random probability-weights $\pi_{jlk}$ satisfy $\sum_{k\ge 1}\pi_{jlk}=1$ a.s. with
\begin{equation}
\label{probweights1}
\pi_{jlk}=\l_{jl}\,(1-\l_{jl})^{k-1},\,\,\,k\ge 1.
\end{equation}
The $\l_{jl}$'s are random geometric-probabilities with 
$\l_{jl}\ind{\cal B}e(\a_{jl},\b_{jl})$ for fixed hyperparameters $\a_{jl}$ and $\b_{jl}$. 
Then, the nonparametric prior over the $Z_j$ error in (\ref{dynamical_equations})
attains the representation
$$
f_j(z)=\sum_{l=1}^mp_{jl}\,M_{jl}(z),
$$
where each $M_{jl}$ is an infinite mixture of normal zero mean kernels via the random
mixing measure $\G_{jl}$ i.e.
$$
M_{jl}(z)=\int_{v>0}{\cal N}(z|0,v^{-1})\,\G_{jl}(dv)=
\sum_{k=1}^\infty\pi_{jlk}\,{\cal N}(z|0,\tau_{jlk}^{-1}).
$$
Clearly, because $\G_{jl}=\G_{lj}$ it is that $M_{jl}=M_{lj}$.

We have the following:
\begin{enumerate}

\item
The random infinite mixtures
$\{M_{jl}:1\le j<l\le m\}$ a-posteriori given the observed time-series, will
capture common characteristics among the pairs of noise densities 
$\{(f_j,f_l):1\le j<l\le m\}$. 

\item
The mixtures $\{M_{jj}:1\le j\le m\}$
a-posteriori will be describing idiosyncratic characteristics of the noise densities 
$\{f_j:1\le j\le m\}$.

\end{enumerate}

\smallskip  
It follows that the model $({\bl x}|{\bl x}_0)$ of the time-series observations  
conditional on the unknown initial conditions, in a hierarchical fashion, is given by
\begin{align}
\label{hierarc1}
& x_{ji}|\,{\bl x}_{j,i:\rho_j},\t_j,\tau_{ji}
  \ind{\cal N}(g_j(\t_j,{\bl x}_{j,i:\rho_j}),\tau_{ji}^{-1})\\
 &\qquad\qquad\text{where  }\,\,1\le i\le n_j,\,1\leq j\leq m\nonumber\\
& \tau_{ji} \given \Q_j \iid \Q_j\nonumber\\
& \Q_j = \sum_{l=1}^mp_{jl}\G_{jl},\,\,\sum_{l=1}^mp_{jl}=1,\,\,\
  \G_{jl}=\G_{lj}\,\,{\rm a.s.}\nonumber\\
& \G_{jl}\ind{\cal GSB}(\a_{jl},\b_{jl},G_0).\nonumber
\end{align}

While our method for pairwise dependent joint reconstruction and prediction can be used for 
dynamical systems where each state $x_{ji}$ depends on the previous $\rho_j$ states ${\bl x}_{j,i:\rho_j}$, 
for simplicity and ease of exposition, in the sequel we will focus on the special case $\rho_j = 1$ for all $j=1,\ldots,m$. Also, with ${\bl x}'$ we will denote the future unobserved observations
along the $m$ multiple time-series, and with ${\bl x}_j'=(x_{j,n_j+1},\ldots,x_{j,n_j+T_j})$ the $T_j$ 
future unobserved observations of the $j$-th time-series.

\subsection{The nonparametric posterior}

Using Bayes' theorem, it is that
\begin{equation}
\label{bayes1}
\Pi({\bl f}, {\bl\t}, {\bl x}_0,{\bl x}'|{\bl x})\propto\Pi({\bl f}, {\bl\t}, {\bl x}_0)\,
\Pi({\bl x}',{\bl x}|{\bl f}, {\bl\t}, {\bl x}_0),
\end{equation}
where $\Pi({\bl f}, {\bl\t}, {\bl x}_0)$ is the prior density over the unknown error processes
$\bl f$, the control parameters $\bl\t$, and the initial conditions ${\bl x}_0$.  
We define the random set $\cal R$ that contains the selection-probabilities ${\bl p}=\{p_{jl}:
1\le j,l\le m\}$, the geometric-probabilities ${\bl\l}=\{\l_{jl}:1\le j\le l\le m\}$ and the 
infinite sequences of the locations 
${\bl\tau}^\infty=\{\tau_{jl}^\infty=(\tau_{jlk})_{k\ge 1}:1\le j\le l\le m\}$
of the GSB random measures $\G_{jl}$. Clearly, we can represent $\cal R$ as the union of 
${\cal R}_j$'s for $j=1,\ldots,m$, with ${\cal R}_j=\{ {\bl p}_j,{\bl\l}_j,{\bl\tau}_j^\infty\}$ 
and ${\bl p}_j=(p_{j1},\ldots,p_{jm})$, ${\bl\l}_j=(\l_{j1},\ldots,\l_{jm})$, and 
${\bl\tau}_j^\infty=(\tau_{j1}^\infty,\ldots,\tau_{jm}^\infty)$. 
Because the estimation of the noise density $\bl f$ is equivalent
to the estimation of the variables in $\cal R$, the right hand side of equation (\ref{bayes1}) 
becomes
$$
\Pi({\cal R},{\bl\t},{\bl x}_0)\prod_{j=1}^m\prod_{i=1}^{n_j+T_j}
 \Pi(x_{ji}|{\cal R}_j, \t_j, x_{j0}),
$$
with the density $\Pi(x_{ji}|{\cal R}_j, \t_j, x_{j0})$ given by
\begin{equation}
\label{random1}
\sum_{l=1}^mp_{jl}\sum_{k=1}^\infty\pi_{jlk}\,
{\cal N}(x_{ji}\,|\,g_j(\t_j,x_{j,i-1}),\tau_{jlk}^{-1}).
\end{equation}
For a finite dimensional Gibbs sampler, we will augment the random densities $f_j$, 
with the following sets of variables for $1\le i\le n_j+T_j$ and $1\le j\le m$:

\begin{enumerate}

\item The GSB-mixture selection variables ${\boldsymbol\delta}=(\d_{ji})$; 
      for an observation $x_{ji}$ that comes from $f_j$, $\d_{ji}$ selects 
      the specific GSB-mixture $M_{ji}$ that the observation came from. 
      It is that P$\{\delta_{ji}=l\}=p_{jl}$.
      
\item The geometric-slice variables ${\mathbf N}=(N_{ji})$, such that $(N_{ji}|\d_{ji}=l)$ 
      follows the negative binomial distribution 
      ${\cal NB}(2,\l_{jl})$, with P$\{N_{ji}=r|\d_{ji}=l\}=r\l_{jl}^2(1-\l_{jl})^{r-1}$
      for all $r\ge 1$.      
        
\item The clustering variables ${\boldsymbol d}=(d_{ji})$; for an observation 
      $x_{ji}$ that comes from $f_j$, given $\delta_{ji}$, $d_{ji}$ allocates 
      the component of the GSB-mixture $M_{j \delta_{ji}}$ that $x_{ji}$ 
      came from. Also, given $N_{ji}$ the variable $d_{ji}$ follows a 
      discrete uniform distribution over the random set ${\cal S}_{ji}=\{1,\ldots,N_{ji}\}$. 
\end{enumerate}
Then the augmented Gibbs sampler will have a dimension of order
$\max\{N_{ji}:1\le j\le m,1\le i\le n_j+T_j\}<\infty$. 

We have the following proposition:

\smallskip\noindent{\bf Proposition 1.} {\sl
Augmenting the random densities given in (\ref{random1}) 
with $(N_{ji},d_{ji},\d_{ji})$ we have
\begin{align}
\label{augmented_density}
& \Pi(x_{ji},N_{ji}=r,d_{ji}=k,\d_{ji}=l\,|\,{\cal R}_j,x_{j,i-1},\t_j)\\
& = p_{ji}\lambda_{jl}^2 \,(1-\lambda_{jl})^{r-1}
  {\cal N}(x_{ji}\,|\,g_{j}(\t_j,x_{j,i-1}),\tau_{jlk}^{-1})\,{\cal I}(k\le r).\nonumber
\end{align}

\noindent 
The proof is given in the Appendix A.
}

\smallskip

From now on, and until the end of this sub-section, we will leave the auxiliary variables 
$N_{ji}$, $d_{ji}$ and $\d_{ji}$ unspecified; especially for the $\d_{ji}$'s we use the notation
$
\delta_{ji}=(\delta_{ji}^1,\ldots,\delta_{ji}^m)\in\{\vec{e}_1,\ldots,\vec{e}_m\},
$
where $\vec{e}_l$ denotes the usual basis vector having its only nonzero component equal to 1 
at position $l$, and P$\{\delta_{ji}=\vec{e}_l|{\bl p}_j\}=p_{jl}$. In fact $(\d_{ji}|{\bl p}_j)$
follows a generalized Bernoulli distribution in the $m$ outcomes $\{\vec{e}_1,\ldots,\vec{e}_m\}$,
whence
\begin{equation}
\label{generalized}
\Pi({\bl\d}|{\bl p})=\prod_{j=1}^m\prod_{i=1}^{n_j}\prod_{l=1}^mp_{ji}^{\d_{ji}^l}.
\end{equation}

We have the following proposition:

\smallskip\noindent{\bf Proposition 2.} {\sl

\begin{enumerate}

\item  
The likelihood
$\Pi({\bl x}',{\bl x},\,{\bl N},{\bl d}\,|\,{\cal R},{\bl\d},{\bl\t},{\bl x}_0)$
conditionally on ${\bl\d}$, is proportional to the triple product:
\begin{align}
\label{triple1}
  & \prod_{j=1}^m\prod_{1\le i\le n_j+T_j\atop d_{ji};\, d_{ji}\le N_{ji}}\prod_{l=1}^m
    \left\{\lambda_{jl}^2(1-\lambda_{jl})^{N_{ji}-1}\tau_{jl d_{ji}}^{1/2}\right.\\
  & \times\left.\exp\left(-{\tau_{jld_{ji}}\over 2} (x_{ji}-g_j(\t_j,\,x_{j,i-1}))^2
     \right)\right\}^{\delta_{ji}^l}.\nonumber     
\end{align}       

\item 
For the special case of Gaussian noise with common precision $\tau$, the likelihood simplifies to: 
$$ 
\prod_{j=1}^m\prod_{i=1}^{n_j+T_j}\tau^{1/2}
\exp\left\{-{\tau\over 2}(x_{ji}-g_j(\t_j,x_{j,i-1}))^2\right\}.
$$

\end{enumerate}

\noindent 
The proof is given in the Appendix A.

\smallskip
The full conditionals for the PD-GSBR Gibbs sampler are given in Appendix B.

}

\section{Numerical illustrations}

In this section, we will demonstrate the efficiency of the proposed 
PD-GSBR sampler for the cases $m=2,3$. 
Using mixture noise processes, with pairwise common characteristics, 
we will illustrate different scenarios in which, joint reconstruction 
can be beneficial in terms of modeling accuracy  for underrepresented 
time-series for which, the independent nonparametric GSBR reconstruction 
turns out to be problematic.

\medskip\noindent{\bf The synthetic time-series:} 
We will generate observations via non-Gaussian quadratic and cubic
autoregressive processes of order one, with chaotic deterministic parts which are given by
$Q_r(x)=1-q_rx^2$, with $q_r\in\{1.65,1.71,1.75\}$, and   
$C_r(x)= 0.05 + c_rx -0.99x^3$, with $c_r\in\{2.55,2.65\}$, respectively.

In the sequel, we will denote by ${\bl x}\sim{\bl g}+{\bl f}$, the fact that 
the $m$-multiple time-series 
${\bl x}=({\bl x}_1^{n_1},\ldots,{\bl x}_m^{n_m})^T$, 
with respective sample sizes $n_1,\ldots,n_m$, has been generated via the dynamical 
systems with deterministic parts ${\bl g}=(g_1,\ldots,g_m)^T$ and noise processes 
distributed as ${\bl f}=({\bl p}\otimes{\bl M})\cdot{\bl 1}$.

\medskip\noindent{\bf Prior specifications:} 
Attempting a noninformative prior specification 
over the geometric-probabilities, we set $a_{jl}=b_{jl}=0.5$. Then all $\l_{jl}$'s,
a-priori will follow the arcsine density ${\cal B}e(0.5,0.5)$ coinciding with the
associated Jeffrey's prior. Previously, the density of the mean measure $G_0$ has been set to 
$g_0={\cal G}(a,b)$. Here we fix the hyperparameters to $a=b=10^{-3}$.
Then the prior density over the $\tau_{jlk}$'s, will be very close to a 
noninformative scale-invariant prior. On the control parameters, and the initial condition
variables, we assign the noninformative translation-invariant priors $\Pi({\bl\t})\propto 1$ and  
$\Pi({\bl x}_0)\propto 1$, respectively. Although such priors are improper (they do not integrate 
to 1) they lead to proper full conditionals. The hyperparameters ${\bl\a}_j$, of the Dirichlet
priors over the selection-probabilities ${\bl p}_j$, will be defined separately, 
for each numerical example.

We will model the unknown deterministic parts, via
the quintic polynomials $g_j(\t_j,x)=\sum_{r=0}^5\t_{jr}x^r$.
For simplicity, we choose to sample only one out-of-sample point i.e. $T_j=1$. In all cases, we
have ran the PD-GSBR Gibbs sampler for $N=60,000$ iterations after a burn-in period of $20,000$
iterations.


\subsection{Borrowing from a cubic to a quadratic map}
For our first numerical example, we have generated $2$-multiple time-series via 
\begin{equation}
\label{m2a}
\left[\begin{array}{c}
{\bl x}_1^{200}\\
{\bl x}_2^{50}\\
\end{array}\right]\!\!\sim\!\!
\left[\begin{array}{c}
C_1\\
Q_1\\
\end{array}\right]
\!\!+\!\!    \left[\!\left[\begin{matrix} 
        0.25   & 0.75 \\
         1     &  0  \\
     \end{matrix}\right]\!\!
\otimes\!\!  
   \left[\begin{matrix} 
         M_{11}    & M_{12} \\
         M_{12}    &  0  \\
     \end{matrix}\right]\!\right]\!\!\cdot\!{\bl 1}.  
\end{equation}
The first time-series has idiosyncratic noise $M_{11}(z)={\cal N}(z|0,10^{-6})$. The density
$M_{12}(z)=0.6{\cal N}(z|0,\s^2)+0.4{\cal N}(z|0,(10\s)^2)$ with $\s^2=3\times 10^{-3}$ is common
for both time-series. So that the noise components perturbing the first and second time-series has been 
sampled from $Z_{1i}\iid 0.25M_{11}+0.75M_{12}$ and $Z_{2i}\iid M_{12}$, respectively.
In this example as initial conditions we took $x_{10}=x_{20}=1$.

In Fig. 1(a), we depict the perturbed cubic trajectory. It can be seen that the time-series
experiences noise induced jumps approximately from the interval $I_1=[-1.60,-0.10)$, containing a chaotic 
attractor, to the interval $I_2=[-0.10, 1.67]$, containing a chaotic repellor, see
\cite{merkatas2017bayesian}. 
The quadratic dynamical system experiences a noise induced escape. 
In Fig. 1(b), we can see that under the intense perturbations of the $M_{12}$ noise, 
the quadratic trajectory, escapes its deterministic invariant set $\X=[-1.11, 1.11]$,
after the first 46 iterations.

\begin{figure}[H]
\centering
\includegraphics[width=0.8\textwidth]{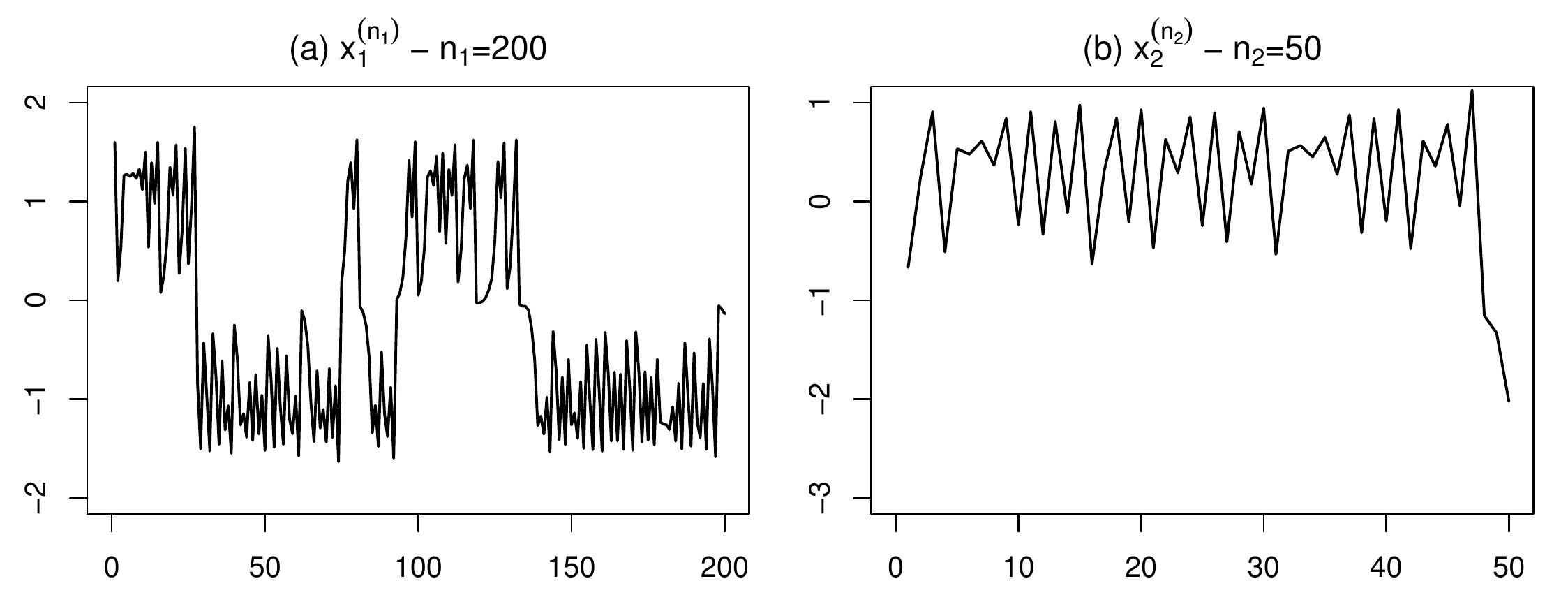}
\caption{The noise perturbed time-series of maps ${\cal C}_1$
         and $Q_1$ are given in Fig. 1(a) and 1(b), respectively.}
\end{figure}

\medskip\noindent{\bf Weak borrowing:} 
To force a weak borrowing scenario a-priori we set 
\begin{equation}
\label{wbor1}
{\bl p}_1\sim{\cal D}ir(10,1)\,\,\,{\rm and}\,\,\,{\bl p}_2\sim{\cal D}ir(1,10),
\end{equation}
with ${\cal D}ir(\a_{j1},\a_{j2})={\cal B}e(\a_{j1},\a_{j2})$. 
For $m=2$ we quantify the borrowing of information (BoI) from the cubic to the quadratic map,
with the posterior mean BoI$_2:=\E(p_{21}|{\bl x})$.
The prior and posterior means of the matrix of the selection-probabilities are given by
$$
\E({\bl p})=
    \left[\begin{matrix} 
        0.909  & 0.091 \\
        0.091  & 0.909
    \end{matrix}\right],\,\,\,
\E({\bl p}|{\bl x})=
    \left[\begin{matrix} 
        0.724   & 0.276 \\
        0.142   & 0.858
    \end{matrix}\right],    
$$
respectively. In this case, the larger data set influences quadratic estimation
by BoI$_2=14.2\%$. 

In Fig. 2(a)-(f), we display in black solid curves the ergodic averages of the 
estimated control parameters, based on 
quintic polynomial modeling, under the weak prior specification in (\ref{wbor1}).
We can see that the ergodic averages, based on the short time-series, given in Fig. 2(g)-(l),
converge to a {\it biased} estimation.

The associated percentage absolute relative errors (PAREs), of the estimated control 
parameters with respect to the true values, are given in the first two lines of Table I.
We can see that the estimations based on the short time-series, exhibit
large errors hindering the identification of map $Q_1$. 

\medskip\noindent{\bf Strong borrowing:} 
To force an a-priori strong borrowing from the map $C_1$ to the map $Q_1$ and
at the same time to be noninformative to the selection-probabilities of $Q_1$, we set
\begin{equation}
\label{sbor1}
{\bl p}_1\sim{\cal D}ir(1,10)\,\,\,{\rm and}\,\,\,{\bl p}_2\sim{\cal D}ir(1,1)={\cal U}(0,1),
\end{equation}
where ${\cal U}(0,1)$, denotes the uniform distribution over the interval $(0,1)$.
We have the following prior and posterior means
$$
\E({\bl p})=
    \left[\begin{matrix} 
        0.091  &  0.909\\
        0.500  &  0.500
    \end{matrix}\right],\,\,\,
\E({\bl p}|{\bl x})=
    \left[\begin{matrix} 
        0.230   & 0.770 \\
        0.927   & 0.073
    \end{matrix}\right].    
$$
The prior specification (\ref{sbor1}) increases borrowing from BoI$_2=14.2\%$ to
$92.7\%$. We remark that the posterior mean of the selection-probabilities for 
the noise process of $Q_1$ in the second row of $\E({\bl p}|{\bl x})$, is much 
closer now to the true selection-probabilities.

In Fig. 2(a)-(f), we can see (in red solid curves) the ergodic 
averages of the control parameters under the strong borrowing scenario.
We can see now that the ergodic averages based on the short time-series, 
given in Fig. 2(g)-(l), are converging fast to the true values.
In the last two lines of Table I, we can see that strong borrowing reduces the average PARE 
of the control parameters of the short time-series, from 2.67\% to a mere 0.37\% 
enabling the identification of the map $Q_1$.

\begin{figure}[H]
\includegraphics[width=0.85\textwidth]{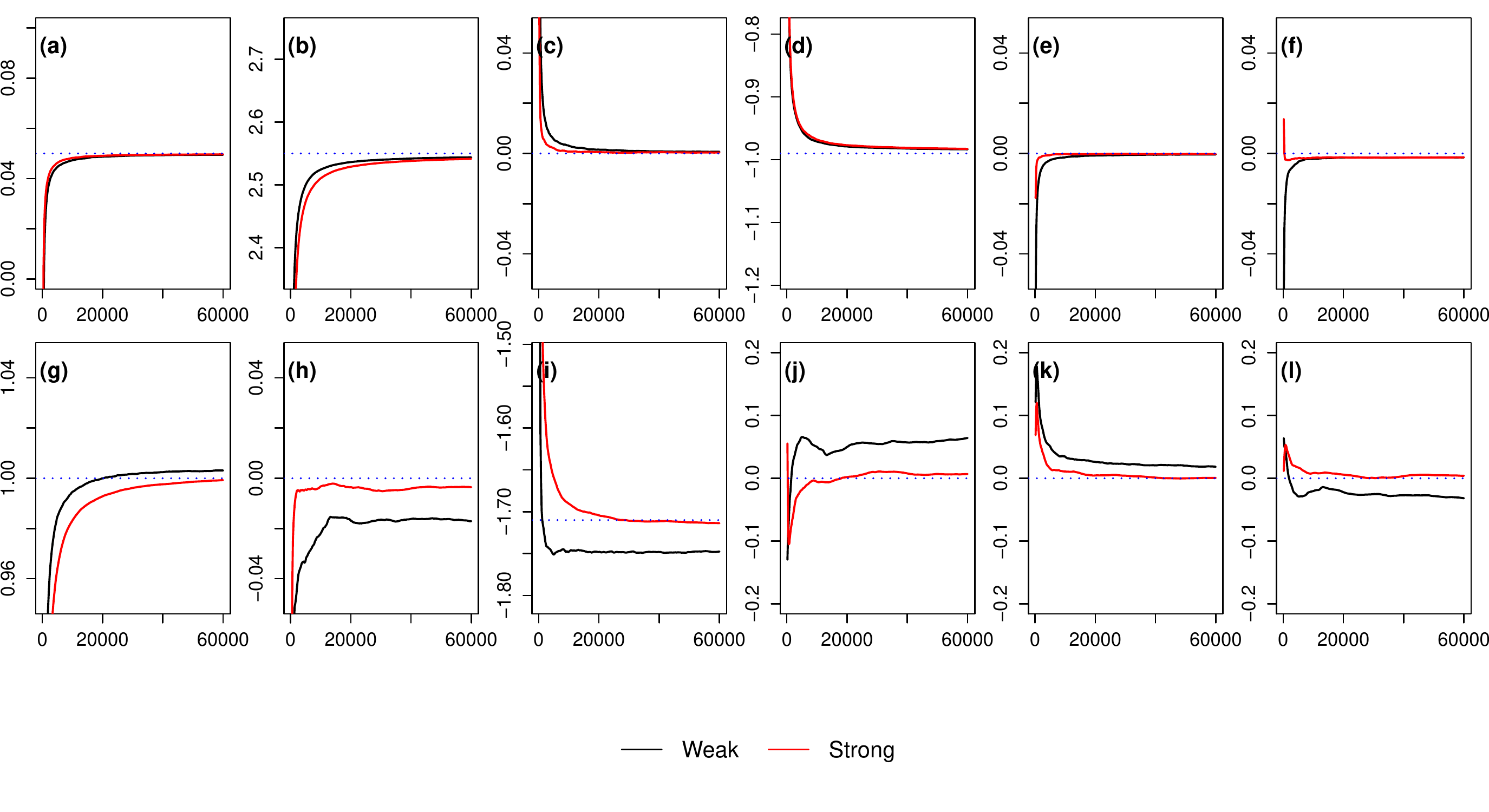}
\centering
\caption{Weak and strong borrowing corresponds to the averages in black and red, respectively. 
         In Fig. 2(a)-(f) and 2(g)-(l) we present the ergodic averages of the control 
         parameters for the maps $C_1$ and $Q_1$, respectively. True control 
         parameter values are represented by blue horizontal dotted lines.}
\end{figure}

\begin{table}[H]
\begin{center}
\caption{PAREs for the PD-GSBR estimation of the control parameters of the $C_1:{\bl x}_1^{200}$ 
         and $Q_1:{\bl x}_2^{50}$ maps.
         The estimation is based on quintic polynomial modeling, under weak and strong borrowing
         priors over the selection-probabilities.}
\begin{tabular}{ccccccccc} 
Borr.       & Map  & $\t_{j0}$ & $\t_{j1}$ & $\t_{j2}$ & $\t_{j3}$ 
            & $\t_{j4}$   & $\t_{j5}$ & $\bar{\t}$  \\
\hline
Weak & $C_1$  & $0.44$ & $0.09$ & $0.04$ & $0.40$ & $0.02$ & $0.14$ & $0.19$  \\ 
     & $Q_1$  & $0.55$ & $1.57$ & $2.39$ & $6.44$ & $1.81$ & $3.24$ & $2.67$ \\ 
\hline
Strong & $C_1$  & $0.50$ & $0.11$ & $0.06$ & $0.50$ & $0.03$ & $0.17$ & $0.23$  \\ 
       & $Q_1$  & $0.23$ & $0.25$ & $0.48$ & $0.57$ & $0.01$ & $0.42$ & $0.37$ \\ 
\hline
\end{tabular}
\end{center}
\end{table}

%
%

\begin{figure}[H]
\includegraphics[width=0.75\textwidth]{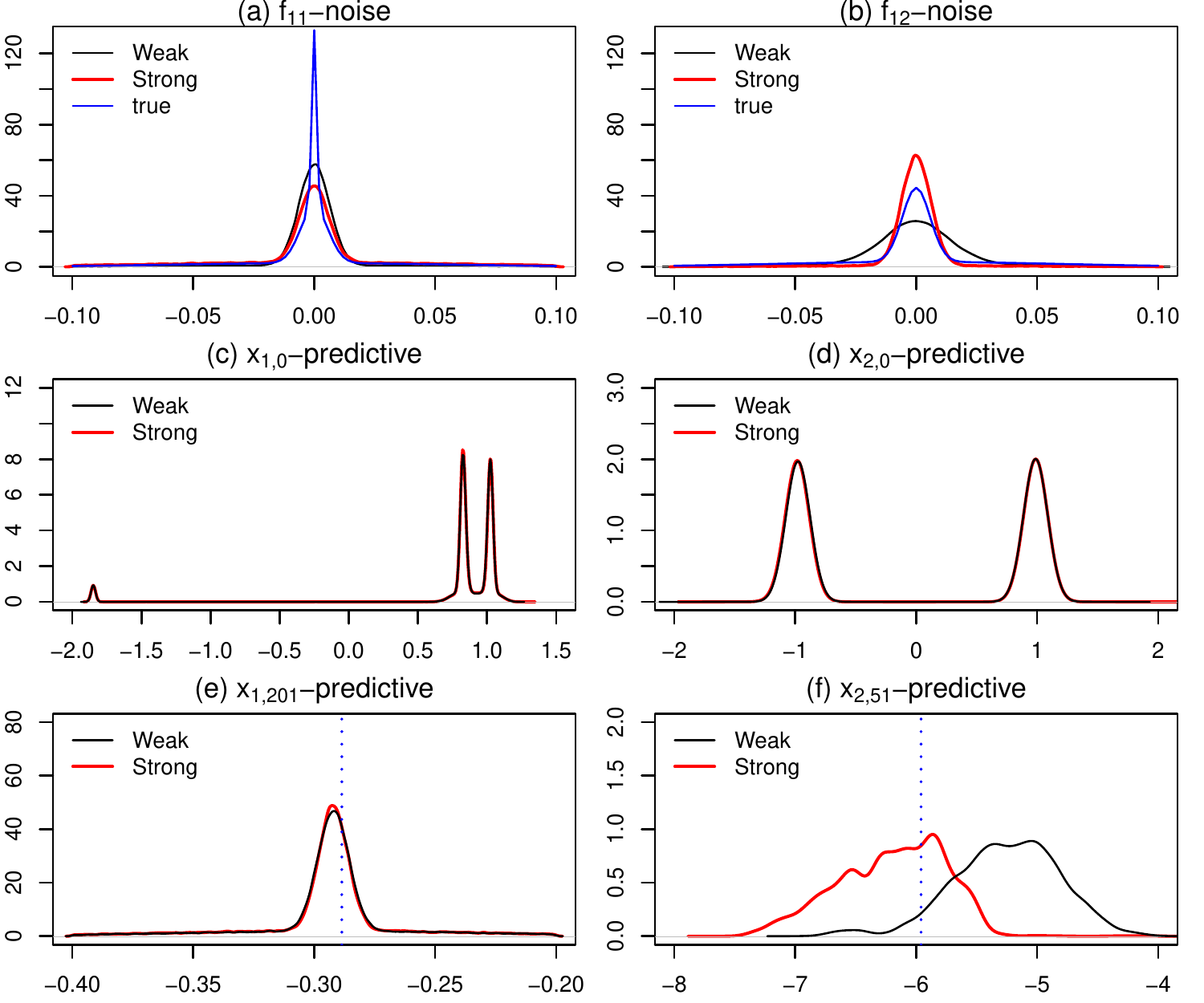}
\centering
\caption{Weak and strong borrowing corresponds 
         to densities in black and red, respectively. 
         Fig. 3(a), 3(c) and 3(e),
         correspond to map $C_1$, and Fig. 3(b), 3(d) and 3(f)
         to the short time-series map $Q_1$. Noise predictive densities
         are given in Fig. 3(a)-(b). Initial conditions predictive densities
         are given in Fig. 3(c)-(d). In Fig. 3(e)-(f), we give the predictive 
         densities of the first future observation. True future values, are
         represented by vertical dotted blue lines.}
\end{figure}

In Fig. 3(a)-(b), we present kernel density estimations (KDEs)  
of the marginal noise densities based on noise predictive samples, 
under weak and strong prior specifications in black and red, respectively.
True noise densities are represented by solid blue curves.

In Fig. 3(c)-(d), we display predictive based KDEs of the marginal posteriors 
of the initial conditions $x_{10}$ and $x_{20}$. The estimations under the two 
prior configurations are nearly indistinguishable.

In Fig. 3(e)-(f), we present predictive based KDEs of the marginal posteriors of the 
out-of-sample variables $x_{1,201}$ and $x_{2,51}$. True future values are represented
in vertical dotted blue lines.
We can see how more accurate is the estimation of the predictive density of the first future 
observation based on the short time-series, lying outside the invariant set, under 
the strong borrowing prior (solid red curve) in Fig. 3(f).


\subsection{Borrowing between two cubic maps}
Here we have generated a pair of cubic time-series via 
\begin{equation}
\label{m1a}
\left[\begin{array}{c}
{\bl x}_1^{200}\\
{\bl x}_2^{30}\\
\end{array}\right]\!\!\sim\!\!
\left[\begin{array}{c}
C_1\\
C_2\\
\end{array}\right]
\!\!+\!\!    \left[\!\left[\begin{matrix} 
        0   & 1\\
        1   & 0\\
     \end{matrix}\right]
\!\!\otimes\!\!  
   \left[\begin{matrix} 
         0       & M_{12}\\
         M_{12}  & 0     \\
     \end{matrix}\right]\!\right]\!\cdot\!{\bl 1}.  
\end{equation}
The mixture 
$M_{12}=0.9{\cal N}(0,\s^2)+0.1{\cal N}(0,(200\s)^2)$ with $\s^2= 10^{-6}$, is playing the
r\^ole of the common noise process.
In this example as initial conditions we took $x_{10}=x_{20}=1$.

The perturbed cubic trajectories are depicted in Fig. 4(a)-(b). It can be seen, that 
both time-series experience noise induced jumps.

\begin{figure}[H]
\includegraphics[width=0.8\textwidth]{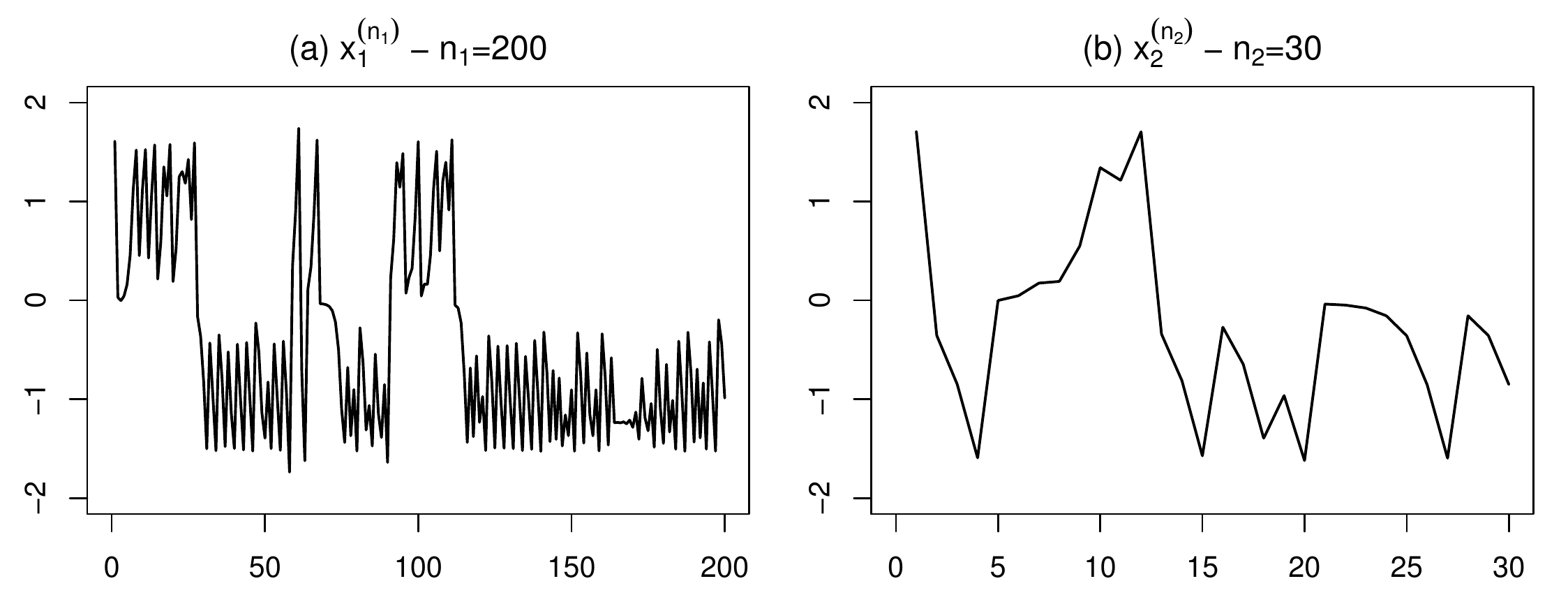}
\centering
\caption{The noise perturbed time-series of maps $C_1$
         and $C_2$, are given in Fig. 4(a) and 4(b), respectively.}
\end{figure}

\medskip\noindent{\bf Weak borrowing:} 
Using the weak prior configuration given in (\ref{wbor1}),
the posterior mean of the matrix of the selection-probabilities is approximated by
$$
\E({\bl p}|{\bl x})=
    \left[\begin{matrix} 
        0.879   & 0.121 \\
        0.100   & 0.900
    \end{matrix}\right].    
$$ 
In this case the large cubic time-series influences the short cubic time-series
by only BoI$_2=10\%$. 

In Fig. 5(a)-(f), we display in solid black curves the ergodic averages of the estimated control 
parameters, under the weak prior specification (\ref{wbor1}).
We can see that the ergodic averages based on the short time-series, given in Fig. 5(g)-(l),
exhibit slow convergence.

The associated PAREs of the estimated control 
parameters with respect to the true values, are given in the first two lines of Table II.
We can see that the estimations based on the short time-series exhibit
large errors, hindering the identification of map $C_2$.

\medskip\noindent{\bf Strong borrowing:} 
To force an a-priori strong borrowing from the map $C_1$ to the map $C_2$, we set
\begin{equation}
\label{sbor2}
{\bl p}_1\sim{\cal D}ir(1,10)\,\,\,{\rm and}\,\,\,{\bl p}_2\sim{\cal D}ir(10,1).
\end{equation}
We have the following prior and posterior means
$$
\E({\bl p})=
    \left[\begin{matrix} 
        0.091  &  0.909\\
        0.909  &  0.091
    \end{matrix}\right],\,\,\,
\E({\bl p}|{\bl x})=
    \left[\begin{matrix} 
        0.005   & 0.995 \\
        0.976   & 0.024
    \end{matrix}\right].    
$$
The prior specification (\ref{sbor2}), increases borrowing considerably from 10\% to
$97.6\%$. We remark that the posterior mean of the selection-probabilities for 
the noise process of $C_2$ in the second row of $\E({\bl p}|{\bl x})$, is close to the true 
selection-probabilities.

In Fig. 5(a)-(f), we can see in solid red curves, the ergodic 
averages of the control parameters under the strong borrowing scenario.
We can see now that the ergodic averages based on the short time-series, 
given in Fig. 5(g)-(l), are converging fast to the true values.
In the last two lines of Table II, we can see that the strong borrowing reduces the average PARE 
of the control parameters of the short cubic time-series, from 1.14\% to a mere 0.10\%, thus 
enabling identification of the map $C_2$.

\begin{figure}[H]
\includegraphics[width=0.85\textwidth]{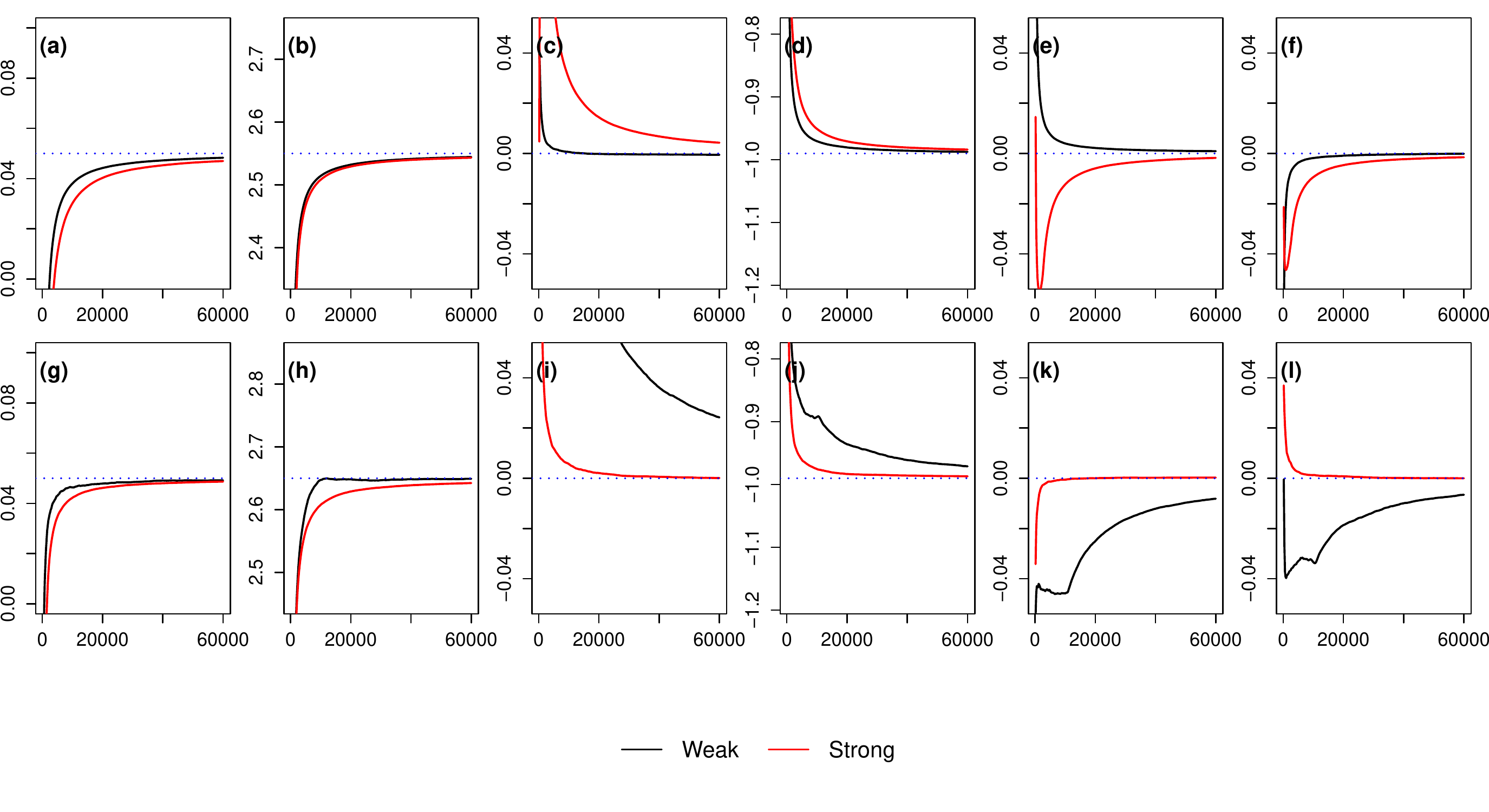}
\centering
\caption{Weak and strong borrowing, corresponds to the averages in black and red, respectively. 
         In Fig. 5(a)-(f) and 5(g)-(l) we present the ergodic averages of the control 
         parameters for the maps $C_1$ and $C_2$, respectively. True control 
         parameter values, are represented by blue horizontal dotted lines.}
\end{figure}

\begin{table}[H]
\begin{center}
\caption{PAREs for the PD-GSBR estimation of the control parameters of the $C_1:{\bl x}_1^{200}$ 
         and $C_2:{\bl x}_2^{30}$ maps.
         The estimation is based on quintic polynomial modeling, under weak and strong borrowing
         priors over the selection-probabilities.}\begin{tabular}{ccccccccc} 
Borr.   & Map & $\t_{j0}$ & $\t_{j1}$ & $\t_{j2}$ & $\t_{j3}$ 
        & $\t_{j4}$  & $\t_{j5}$ & $\bar{\t}$                            \\
\hline
Weak & $C_1$ & $0.36$ & $0.01$ & $0.06$ & $0.00$ & $0.02$ & $0.00$ &  $0.08$\\ 
     & $C_2$ & $1.16$ & $0.19$ & $2.36$ & $1.70$ & $0.78$ & $0.66$ &  $1.14$\\ 
\hline
Strong & $C_1$ & $0.60$ & $0.02$ & $0.50$ & $0.34$ & $0.21$ & $0.16$ & $0.31$\\ 
       & $C_2$ & $0.31$ & $0.03$ & $0.09$ & $0.04$ & $0.04$ & $0.09$ & $0.10$\\ 
\hline
\end{tabular}
\end{center}
\end{table}

%
%

\begin{figure}[H]
\includegraphics[width=0.75\textwidth]{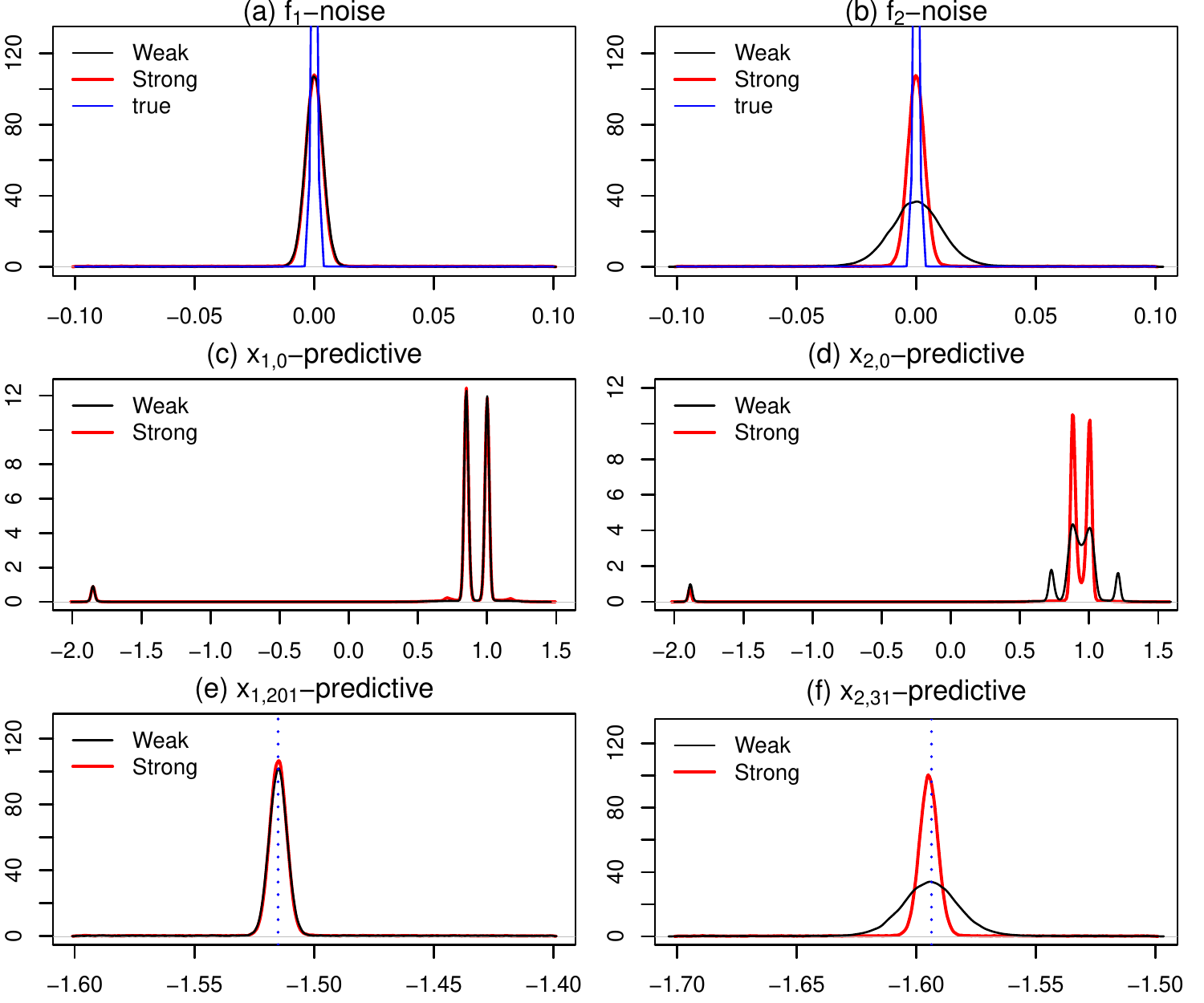}
\centering
\caption{Weak and strong borrowing corresponds 
         to densities in black and red, respectively. 
         Fig. 6(a), 6(c) and 6(e),
         correspond to map $C_1$, and Fig. 6(b), 6(d) and 6(f)
         to the short time-series map $C_2$. Noise predictive densities
         are given in Fig. 6(a)-(b). Initial conditions predictive densities
         are given in Fig. 6(c)-(d). In Fig. 6(e)-(f), we give the predictive 
         densities of the first future observation. True future values are
         represented in vertical dotted blue lines.}
\end{figure}

In Fig. 6(a)-(b), we present the KDEs  
of the marginal noise densities based on the noise predictive samples, 
under weak and strong prior specifications, in black and red solid curves, 
respectively. True noise densities are represented by solid blue curves.
We remark the similarity of the estimated noise densities under the 
strong prior configuration (\ref{sbor2}).

In Fig. 6(c)-(d), we display the predictive based KDEs of the marginal 
posteriors of the initial points $x_{10}$ and $x_{20}$. The estimated marginal 
posterior density of the variable $x_{20}$, under the weak borrowing prior has five modes.
The two spurious modes, disappear after the 
introduction of strong borrowing (solid curve in red). 
We remark that the three modes are very close to the three real roots of the polynomial 
equation $C_2(x)-x=0$.

In Fig. 6(e)-(f), we present the predictive based KDEs of the marginal posteriors of the first 
out-of-sample variables $x_{1,201}$ and $x_{2,31}$. The point estimations of the first out-of-sample 
values are of the same quality, yet,
under strong borrowing the predictive density associated with the short
time-series cubic map exhibits a 95\% highest posterior density interval (HPDI) 
shrinkage factor of 0.45. Namely, the weak borrowing HPDI $(-1.622,-1.566)$ of 
the variable $x_{2,31}$ shrinks to the strong borrowing HPDI $(-1.607,-1.582)$.  

\subsection{Borrowing between three quadratic maps}

For this example we have generated a $3$-multiple perturbed quadratic time-series via
{\small 
\begin{equation}
\label{meq3}
     \left[\!\begin{array}{c}
        {\bl x}_1^{200}\\
        {\bl x}_2^{20} \\
        {\bl x}_3^{200}\\
     \end{array}\!\right]\!\!\sim\!\!
     \left[\!\begin{array}{c}
           Q_3\\
           Q_2\\
           Q_1\\
     \end{array}\!\right]
\!\!+\!\!    \left[\!\left[\begin{matrix} 
        0     &  1      &  0   \\
        0.90  &  0      &  0.10\\
        0     &  0.33   &  0.67\\
     \end{matrix}\right]
\!\!\otimes\!\!  
     \left[\begin{matrix} 
         0       & M_{12} &  0     \\
         M_{12}  & 0      &  M_{23}\\
         0       & M_{23} &  M_{33}\\
     \end{matrix}\right]\!\right]\!\cdot\!{\bl 1}.  
\end{equation}
}
The first two time-series have the common part $M_{12}={\cal N}(0,10^{-6})$ and 
no idiosyncratic parts. The second and third time-series have the common part 
$M_{23}={\cal N}(0,4\times10^{-2})$. The third time-series has idiosyncratic
part $M_{33}={\cal N}(0,10^{-4})$.
In this example as initial conditions we took $x_{10}=x_{20}=1$ and $x_{30}=0.5$.

The three perturbed quadratic trajectories are displayed in Fig. 7(a)-(c).

\begin{figure}[H]
\includegraphics[width=0.45\textwidth]{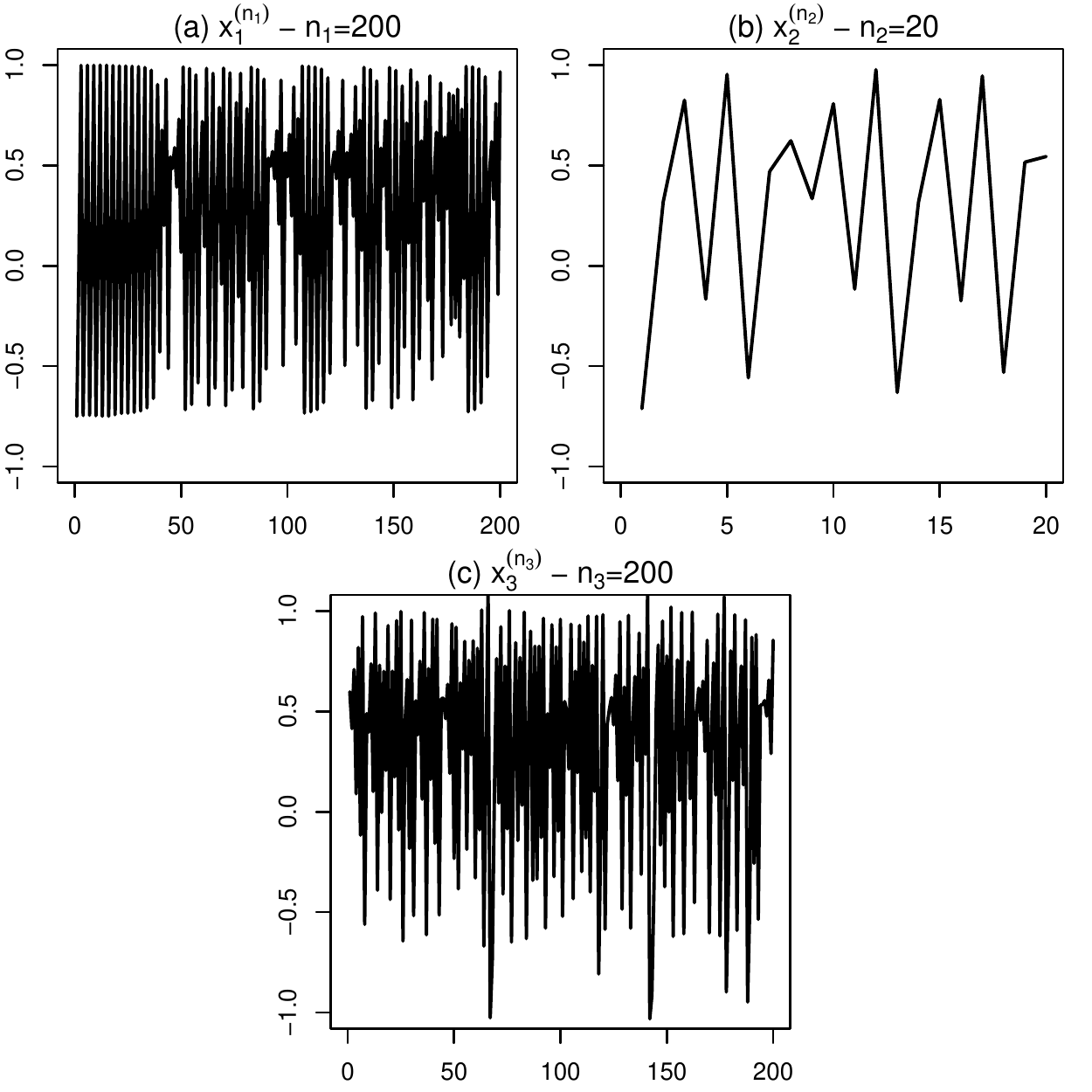}
\centering
\caption{The noise perturbed time-series for the maps $Q_3$, $Q_2$
         and $Q_1$, are given in Fig. 7(a), 7(b) and 7(c), respectively.}
\end{figure}

\medskip\noindent{\bf Weak borrowing:} 
To force a weak borrowing scenario, a-priori we set for $j=1,2,3$
\begin{equation}
\label{wbor3}
{\bl p}_j\sim{\cal D}ir({\bl\a}_j),\,\,\a_{ji}=10\,{\cal I}(j=i)+{\cal I}(j\neq i),
\end{equation}
with prior mean $\E({\bl p})=10/12\,{\cal I}(j=i)+1/12\,{\cal I}(j\neq i)$.
Then the posterior mean after stationarity, is approximated by
$$
\E({\bl p}|{\bl x})=
     \left[\begin{matrix}
         0.990 & 0.005 & 0.005\\
         0.032 & 0.912 & 0.056\\
         0.105 & 0.207 & 0.688
     \end{matrix}\right].
$$
For $m=3$, we quantify the borrowing of information from the first and third time-series
to the second, with the posterior mean BoI$_3:=\E(p_{21}+p_{23}|{\bl x})$.
Because, BoI$_3=0.88\%$, the two large time-series have a very small
effect on the central short time-series.

In Fig. 8(a)-(r), we display in solid black curves the ergodic averages of the estimated control 
parameters, under the weak prior specification (\ref{wbor3}).
We can see that the ergodic chains, based on the short time-series, Fig. 8(g)-(l),
exhibit serious {\it mixing issues}.

The associated PAREs of the estimated control 
parameters with respect to the true values, are given in the first three lines of Table III.
We can see that the estimations based on the short time-series ${\bl x}_2^{20}$, exhibit
large errors, hindering the identification of the map $Q_2$. 
This situation can be corrected by the introduction of a strong borrowing prior configuration.

\medskip\noindent{\bf Strong borrowing:} 
To force an a-priori strong borrowing from the maps $Q_3$ and $Q_1$ to the map $Q_2$, and
at the same time to be noninformative on the selection-probabilities of $Q_2$, we set for $j=1,2,3$
\begin{equation}
\label{sbor3}
{\bl p}_j\sim{\cal D}ir({\bl\a}_j),\,\,\a_{ji}=10\,{\cal I}((j,i)\in J)+{\cal I}((j,i)\notin J),
\end{equation}
where $J=\{(j,i):j\neq 2, i=2\}$.
We have the following prior and posterior means
$$
\E({\bl p})=
    \left[\begin{matrix} 
        0.083  &  0.834  & 0.083 \\
        0.333  &  0.334  & 0.333 \\
        0.083  &  0.834  & 0.083
    \end{matrix}\right],
$$
and
$$    
\E({\bl p}|{\bl x})=
    \left[\begin{matrix}
       0.005 & 0.990 & 0.005\\
       0.849 & 0.045 & 0.106\\
       0.125 & 0.692 & 0.183
    \end{matrix}\right].    
$$
The prior specification (\ref{sbor3}) increases borrowing from BoI$_3=0.88\%$ to
$95.5\%$. We remark how close is the posterior mean of the selection-probabilities for 
the noise process of $Q_2$, in the second row of $\E({\bl p}|{\bl x})$, to the true 
selection- probabilities in (\ref{meq3}).

In Fig. 8(a)-(r), we can see in solid red curves the ergodic 
averages of the control parameters, under the strong borrowing scenario.
We can see now that the ergodic averages based on the short time-series, given in Fig. 8(g)-(l), 
are converging fast to the true values.
In the last three lines of Table III, we can see that strong borrowing reduces the average PARE 
of the control parameters of the short time-series, from 12.87\% to a mere 0.17\% 
enabling the identification of the map $Q_2$.

\begin{figure}[H]
\includegraphics[width=0.85\textwidth]{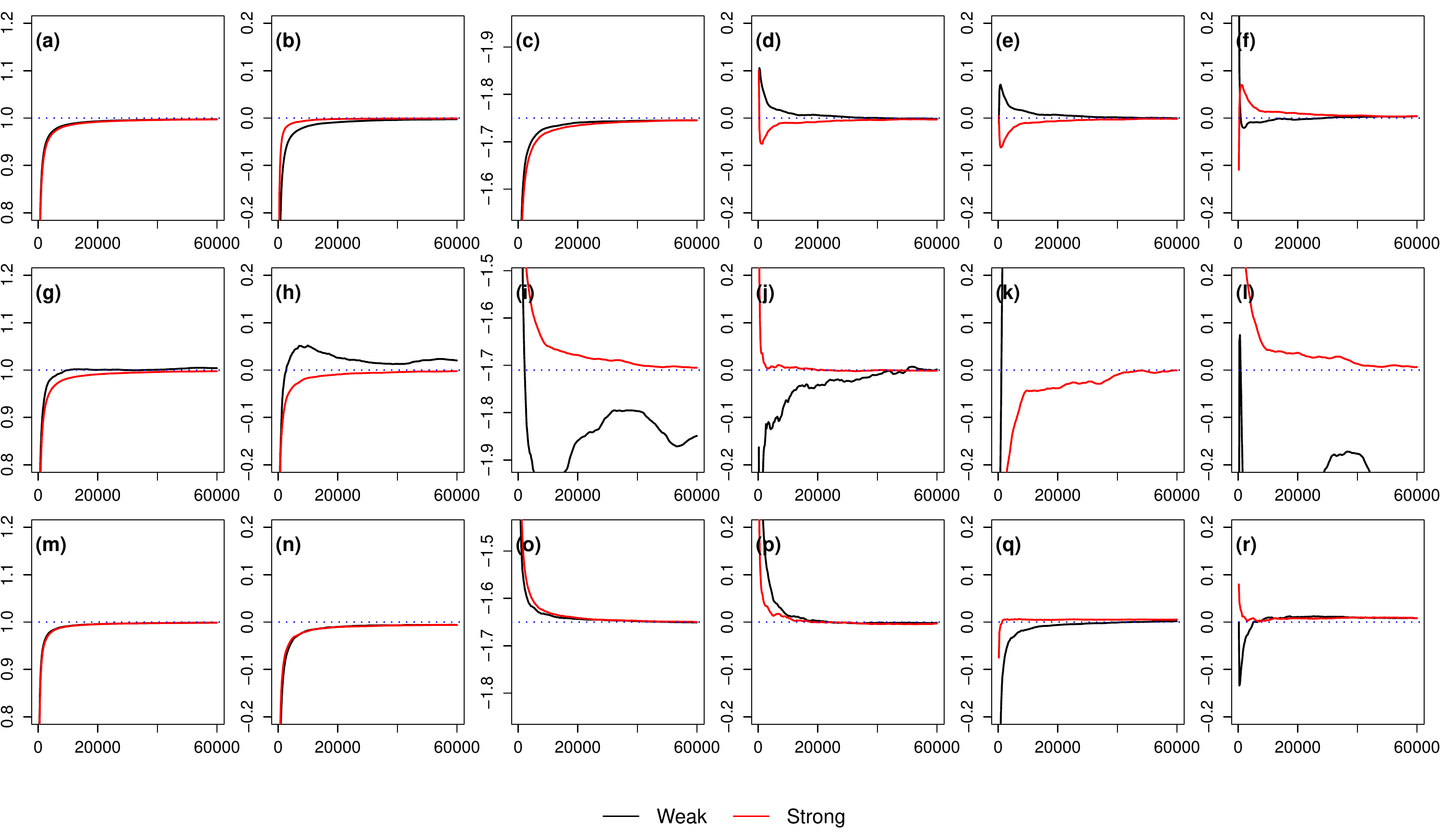}
\centering
\caption{Weak and strong borrowing corresponds to the averages in black and red solid curves,
         respectively. 
         In Fig. 8(a)-(f), 8(g)-(l) and 8(m)-(r) we present the ergodic averages of the control 
         parameters for the maps $Q_3$, $Q_2$ and $Q_1$, respectively.  True control 
         parameter values are represented by horizontal dotted blue lines.}
\end{figure}

\begin{table}[!htb]
\begin{center}
\caption{PAREs for the PD-GSBR estimation of the control parameters of the $Q_3:{\bl x}_1^{200}$, 
         $Q_2:{\bl x}_2^{20}$ and $Q_1:{\bl x}_2^{200}$ maps.
         The estimation is based on quintic polynomial modeling, under weak and strong borrowing
         priors over the selection-probabilities.}
\begin{tabular}{ccccccccc} 
Borr.        & Map  & $\t_{j0}$ & $\t_{j1}$ & $\t_{j2}$ & $\t_{j3}$ 
             &      $\t_{j4}$   & $\t_{j5}$ & $\bar{\t}$  \\
\hline
Weak         & $Q_3$  & $0.00$ & $0.00$ & $0.02$ & $0.14$ & $0.00$  & $0.22$  & $0.06$  \\ 
             & $Q_2$  & $0.62$ & $2.38$ & $8.51$ & $0.33$ & $38.17$ & $27.19$ & $12.87$ \\ 
             & $Q_1$  & $0.02$ & $0.35$ & $0.25$ & $0.49$ & $0.54$  & $0.84$  & $0.41$  \\ 

\hline
Strong       & $Q_3$  & $0.02$ & $0.05$ & $0.04$ & $0.37$ & $0.19$ & $0.49$ & $0.19$ \\ 
             & $Q_2$  & $0.05$ & $0.01$ & $0.11$ & $0.27$ & $0.17$ & $0.39$ & $0.17$ \\ 
             & $Q_1$  & $0.03$ & $0.35$ & $0.27$ & $0.54$ & $0.56$ & $0.87$ & $0.43$ \\ 

\hline
\end{tabular}
\end{center}
\end{table}

In Fig. 9(a)-(c), we display the KDEs based on the marginal noise predictive samples. 
In Fig. 9(d)-(f), we display the KDEs based on the marginal initial conditions variable
samples. In Fig. 9(g)-(i), we exhibit the KDEs based on 
the marginal posterior samples of the first out-of-sample variables.
Black and red solid curves, refer to weak and strong borrowing priors, respectively. 
In Fig. 9(b), we have superimposed the noise predictives coming from the weak and strong
borrowing scenarios together with the true density of the noise component, given by $0.9M_{12}+0.1M_{23}$,
in black, red and blue solid curves, respectively.
We note how close to the true noise density, is the density estimated under strong borrowing.
In Fig. 9(e), the KDE based on the marginal posterior predictive of the initial condition
sample under the strong borrowing prior, has its modes very close to $-1$ and 1. In Fig. 9(h),
the estimation of the first out-of-sample value under the strong borrowing prior, exhibits a
shrinked 95\% HPDI at $(0.484,0.503)$.

\begin{figure}[H]
\includegraphics[width=0.75\textwidth]{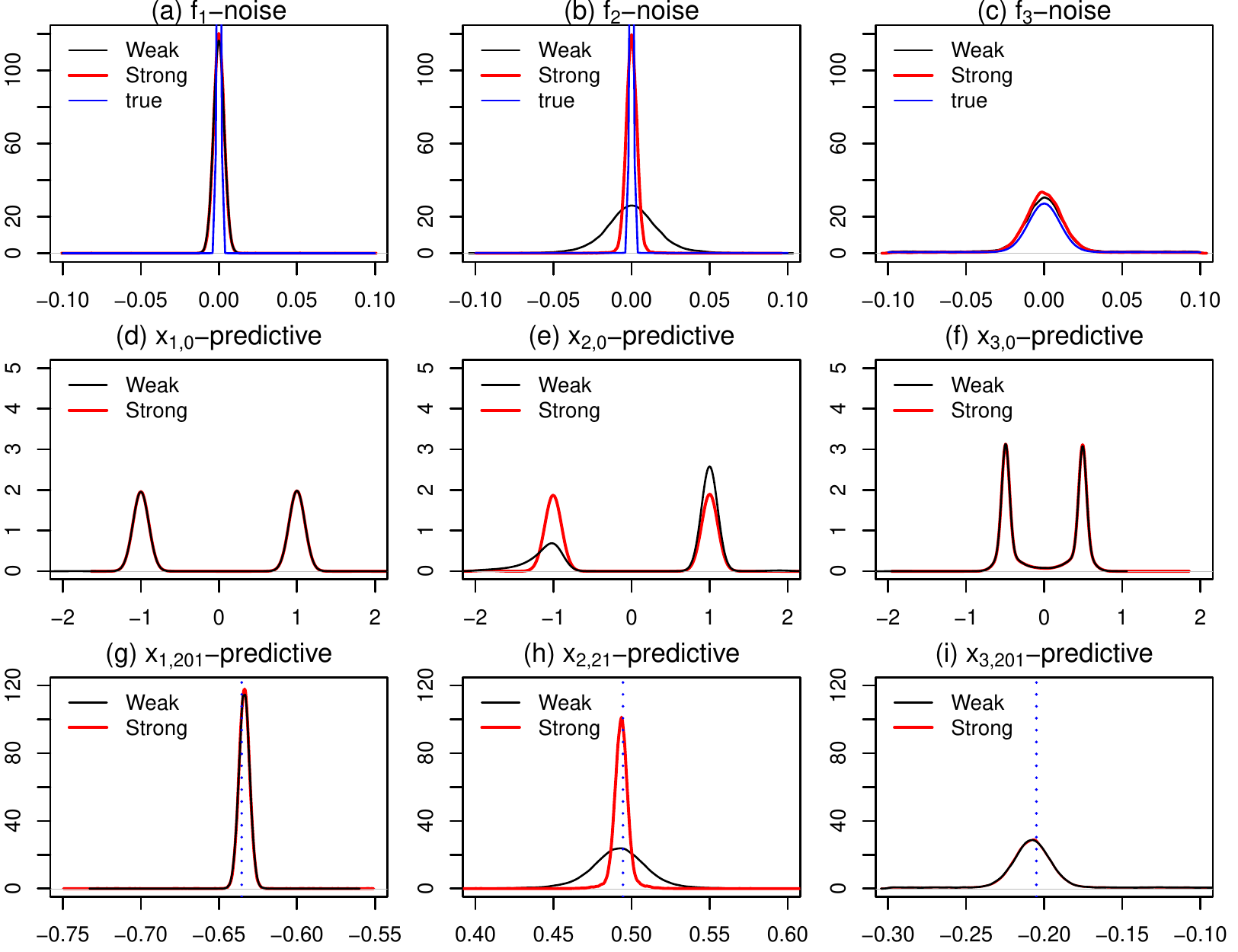}
\centering
\caption{Weak and strong borrowing corresponds to densities in black and red, 
         respectively. Fig. 9(a), 9(d) and 9(g),
         correspond to map $C_3$, Fig. 9(b), 9(e) and 9(h)
         to the short time-series map $C_2$, and Fig. 9(c), 9(f) and 9(i) to map $C_1$
         Noise predictive densities are given in Fig. 9(a)-(c). 
         Initial conditions predictive densities
         are given in Fig. 9(d)-(f). In Fig. 9(g)-(i), we display the predictive 
         densities of the first future observation. True future values, are
         represented by vertical dotted blue lines.}
\end{figure}

\section{Conclusions}

We have proposed a new Bayesian nonparametric model, for the joint pairwise dependent reconstruction 
of dynamical equations, based on observed chaotic time-series data contaminated by dynamical
noise. Also, we have introduced a joint parametric Gibbs sampler. In this case the dynamical noise is 
assumed to be Gaussian, coming from the same noise source for each time-series. 
Then borrowing of strength, comes from the full conditional of the common precision.

Our numerical experiments, are indicating, that when the densities of the noise
processes have common characteristics, underrepresented time series for which
an independent Bayesian nonparametric estimation is problematic, can benefit in 
terms of model estimation accuracy. This can be done by imposing strong borrowing 
prior specifications between the selection-probabilities of the noise processes 
of the short time-series, and the time-series with an adequate 
number of observations for independent Bayesian nonparametric estimation.

Our model can be generalized to include all possible dependencies between the components of the noise
processes. For example, consider the set $[m]_j$ of the first $m$ natural numbers, except $j$, for $1\le j\le m$.
We define the set, $C_{mjl}$, of combinations without replacement, over the set of symbols $[m]_j$,
$l$ at a time, with $0\le l\le m-1$, and $C_{mj0}=\emptyset$. Now to each combination 
$\eta\in C_{mjl}$, we add the symbol $j$, and order the resulting sequence of numbers to $\eta^*$, 
we set $C_{mjl}^*=\{\eta^*:\eta\in C_{mjl}\}$. The set $C_{mjl}^*$ contains the indexes of all
possible interactions of the noise process $f_j$ of order $l+1$. Then the nonparametric prior 
over the $j$th noise process can be written as $f_j(z)=\int_{v>0}{\cal N}(z|0,v^{-1})\Q_j(dv)$, with
$$
\Q_j=\sum_{l=0}^{m-1}\sum_{\xi\in C_{mjl}^*}p_{j,\xi}\,\G_\xi,\quad
\G_\xi\ind{\cal GSB}(\a_\xi,\b_\xi,G_0).
$$
and $\sum_{l=0}^{m-1}\sum_{\xi\in C_{mjl}^*}p_{j,\xi}=1$ a.s. Then the $f_j$ density, will be the  
random mixture of the $2^{m-1}$ GSB random mixtures $M_\xi$ for $\xi\in C_{mjl}^*$. 
The total number of the independent GSB processes needed 
to model $\bl f$, will be $2^m-1$.

\appendix

\section{Proofs of propositions}

\medskip\noindent{\bf Proof of Proposition 1.}
Augmenting the random densities given in (\ref{random1}) with $N_{ji}$ we have: 
{\small
\begin{eqnarray}
&  &    \Pi(x_{ji},N_{ji}=r\,|\,{\cal R}_j, x_{j,i-1},\t_j) 
      = \sum_{l=1}^{m} \Pi(x_{ji},N_{ji}=r,\delta_{ji}=l\,|\,{\cal R}_j,x_{j,i-1},\t_j) \nonumber\\
&  &  = \sum_{l=1}^{m} \Pi(\delta_{ji}=l)\sum_{k=1}^{\infty} 
        \Pi(x_{ji},N_{ji}=r,d_{ji}=k\,|\,{\cal R}_j,x_{j,i-1},\delta_{ji}=l,\t_j)\nonumber\\
&  &  = \sum_{l=1}^{m}p_{jl}\sum_{k=1}^{\infty}   
        \Pi(N_{ji}=r\,|\,{\cal R}_j,\delta_{ji}=l)\Pi(d_{ji}=k\,|\,N_{ji}=r)
        \Pi(x_{ji}\,|\,x_{j,i-1},d_{ji}=k,\delta_{ji}=l,\t_j)\nonumber\\
&  &  = \sum_{l=1}^{m} p_{jl}\sum_{k=1}^{\infty}
        {\cal NB}(N_{ji}=r\given 2,\lambda_{jl})\,
        {\cal DU}(k| 1,r)\,
        {\cal N}(x_{ji}\,|\,g_{j}(\theta_{j},x_{j,i-1}),\tau_{jlk}^{-1})\nonumber\\
&  &  = \sum_{l=1}^mp_{jl}\sum_{k=1}^r
        \l_{jl}^2(1-\l_{jl})^{r-1}\,{\cal N}(x_{ji}\,|\,  
        g_j(\t_j,x_{j,i-1}),\tau_{jlk}^{-1}),\nonumber
\end{eqnarray}
}
where ${\cal DU}(k|1,r)=r^{-1}{\cal I}(k\le r)$ is the discrete uniform distribution,
over the set $\{1\ldots,r\}$. Then, it is clear that the $(N_{ji},d_{ji},\d_{ji})$-augmented 
density is given by (\ref{augmented_density}).\hfill $\square$


\medskip\noindent{\bf Proof of Proposition 2.}\\
\smallskip\noindent{1.}
From equation (\ref{augmented_density}), in vector notation for $\d_{ji}$, it is that
\begin{align}
\Pi & (x_{ji},N_{ji},d_{ji}|{\cal R}_j,\d_{ji},x_{j,i-1},\t_j) = {\cal I}(d_{ji}\le N_{ji})\nonumber\\
    & \times\prod_{l=1}^m\left\{\l_{jl}^2 \,(1-\l_{jl})^{N_{ji}-1}
     {\cal N}(x_{ji}\,|\,g_j(\t_j,x_{j,i-1}),\tau_{jld_{ji}}^{-1})\right\}^{\d_{ji}^l}.\nonumber
\end{align}
The desired result comes from the substitution of the last equation in the 
conditional likelihood expression 
\begin{align}
\Pi & ({\bl x}',{\bl x},\,{\bl N},{\bl d}\,|\,{\cal R},{\bl\d},{\bl\t},{\bl x}_0)\nonumber\\
    & =\prod_{j=1}^m\prod_{i=1}^{n_j+T_j}\Pi(x_{ji},N_{ji},d_{ji}\,|\,
      {\cal R}_j,\d_{ji},x_{j,i-1},\t_j).\nonumber
\end{align}

\smallskip\noindent{2.}
Fixing the random selection probabilities to $p_{jl}=1/(m-1)\,{\cal I}(l<j)$ and the 
random mixing measures to $\G_{jl}=\d_\tau$ a.s., it is that
$$
\Pi({\bl x}',{\bl x}\,|\,{\bl\t},{\bl x}_0)
=\prod_{j=1}^m\prod_{i=1}^{n_j+T_j}\Pi(x_{ji}\,|\,x_{j,i-1},\t_j),
$$
with $\Pi(x_{ji}\,|\,x_{j,i-1},\t_j)={\cal N}(x_{ji}\,|\,0,\tau^{-1})$,
which gives the desired result.\hfill $\square$

\section{Full conditional distributions for the PD-GSBR Gibbs sampler}

In this appendix we describe the PD-GSBR Gibbs sampler. 
At each iteration of the Gibbs sampler we will sample the variables:
\begin{align}
\nonumber
   & \tau_{jlk}, 1\le j \le l \le m,\,1\le k \leq N^*,\\
\nonumber
   & N_{ji},d_{ji},\d_{ji}, 1\le j \le m,\,1\le i \le n_j+T_j,\\
\nonumber
   & p_{jl}, 1\leq j \le m, 1\le l \le m,\\
\nonumber
 & \vartheta_j, x_{j0}, 1\leq j \leq m
\end{align}
with $N^*:=\max_{j,i}N_{ji}$ a.s. finite.
Having in mind that for ${\cal R}=\{{\bl p},{\bl\l},{\bl\tau}^\infty\}$
the $({\bl N},{\bl d},\,{\bl \d})$-augmented posterior is proportional to
\begin{align}
\Pi({\bl p})\Pi({\bl\l}) & \Pi({\bl\tau}^\infty)
    \Pi({\bl\t})\Pi({\bl x}_0)\Pi({\bl\d}|{\bl p})\nonumber\\
   & \times\Pi({\bl x}',{\bl x},\,{\bl N},{\bl d}\,|\,{\bl p},
     {\bl\l},{\bl\tau}^\infty,{\bl\d},{\bl\t},{\bl x}_0),\nonumber    
\end{align}
and taking into account (\ref{generalized}) and (\ref{triple1})
we have the following:

\smallskip
\noindent {\bf 1.}
Letting ${\cal H}_{jilk}:={\cal N}(x_{ji}| g_j(\t_j,x_{j,i-1}),\tau_{jlk}^{-1})$,
and ${\cal I}_{jilk}:={\cal I}(\d_{ji}=\vec{e}_l,d_{ji}=k)$,
the full conditionals for the precisions $\tau_{jlk}$, for $k=1,\ldots,N^*$
and $1\le j\le l\le m$, are given by
\begin{equation}
\label{precis1}
\Pi(\tau_{jlk}|\cdots) \propto\, 
\Pi(\tau_{jlk})\prod_{i=1}^{n_j}{\cal H}_{jilk}^{{\cal I}_{jilk}}
\prod_{i=1}^{n_l}{\cal H}_{jilk}^{{\cal I}(j<l){\cal I}_{lijk}},
\end{equation}
where $\Pi(\tau_{jlk}|\cdots)$ denotes the dependence of the variable $\tau_{jlk}$
to the rest of the variables. 
Standard Bayesian modeling suggests, the use of gamma conjugate prior distributions over the
$\tau_{jlk}$'s, so, we set $\tau_{jlk}\sim g_0={\cal G}(a,b)$, where $g_0$ stands for 
the density of the mean measure $G_0$ which is
a gamma density with shape $a$, rate $b$. 
Then, letting $h_{\t_j}(x_{ji},x_{j,i-1}) := (x_{ji}-g_j(\t_{j},x_{j,i-1}))^2$, it is not difficult to verify that the full conditional of $\tau_{jlk}$ is gamma with shape
$$
a+{1\over 2}\sum_{i=1}^{n_j}{\cal I}_{jilk}+
{1\over 2}{\cal I}(j<l)\sum_{i=1}^{n_l}{\cal I}_{lijk},
$$
and rate
\begin{align}
b & +{1\over 2}\sum_{i=1}^{n_j}{\cal I}_{jilk}\,h_{\t_j}(x_{ji},x_{j,i-1})\nonumber\\
  & +{1\over 2}{\cal I}(j<l)\sum_{i=1}^{n_l}\,{\cal I}_{lijk}\,h_{\t_l}(x_{li},x_{l,i-1}).\nonumber
\end{align}

\smallskip
\noindent {\bf 2}. 
Next, we will sample the mixture allocation variables $d_{ji}$ and the mixture 
component indicator variables $\delta_{ji}$ as a {\it block}. 
For $j=1,\ldots,m$ and $i=1,\ldots,n_j+T_j,$ it is that
$$
{\rm P}\{d_{ji}=k,\d_{ji}=\vec{e}_l|\cdots\}\propto
 p_{jl}\,{\cal H}_{jilk}\,{\cal I}(l\le m)\,{\cal I}(k\le N_{ji}).
$$

\medskip
\noindent {\bf 3.} 
The geometric-slice variables $N_{ji}$ have full conditional distributions 
that are given by
$$
{\rm P}\{N_{ji} = r|\d_{ji}=\vec{e}_l,d_{ji}=l,\cdots\}
\propto(1-\lambda_{jl})^r\,{\cal I}(l\leq r),
$$
which are truncated geometric distributions over the set $\{l, l+1,\ldots\}$.

\medskip
\noindent {\bf 4.} 
The full conditional for the selection-probabilities ${\bl p}_j$, $j=1,\ldots,m$, 
under the conjugate Dirichlet prior 
$$
\Pi({\bl p}_j)={\cal D}ir({\bl p}_j|{\bl\a}_j)
={\Gamma(\a_{j1}+\cdots+\a_{jm})\over
\Gamma(\a_{j1})\cdots\Gamma(\a_{jm})}
\prod_{l=1}^m p_{jl}^{\a_{jl}-1},
$$
with fixed hyperparameter
${\bl\a}_j=(\a_{j1},\dots,\a_{jm})$, supported over the probability simplex
$\{{\bl p}_j\in(0,1)^m:\sum_{l=1}^mp_{jl}=1\}$, is the Dirichlet distribution
$$
\Pi({\bl p}_j|\cdots)\,\propto\,\prod_{l=1}^m p_{jl}^{\a_{jl}+\sum_{i=1}^{n_l}{\cal I}(\delta_{ji}\,=\,\vec{e}_l) - 1}.
$$

\medskip
\noindent {\bf 5.} 
The full conditionals for the geometric-probabilities $\l_{jl}$'s, 
under beta conjugate priors $\l_{jl}\sim{\cal B}e(a_{jl},b_{jl})$, 
are beta distributions. Defining
$S_{jl}: =\sum_{i=1}^{n_j}{\cal I}(\d_{ji}={\bf e}_l)$ and 
$S_{jl}':=\sum_{i=1}^{n_j}{\cal I}(\d_{ji}={\bf e}_l)(N_{ji}-1)$, it is that
\begin{align}
f(\l_{jl}|\cdots) = {\cal B}e(\l_{jl}|
a_{jl} & + 2(S_{jl}+S_{lj}{\cal I}(l<j)),\nonumber\\
 & b_{jl} + (S_{jl}' + S_{lj}'{\cal I}(l<j))).\nonumber
\end{align}

\medskip
\noindent {\bf 6.} 
For the vectors of control parameters $\t_j$ $1\leq j \leq m$ 
the full conditional becomes
\begin{equation}
\label{fullcondthetajoint}
\Pi(\t_j|\cdots)\propto\Pi(\t_j)
\exp\{-\frac{1}{2}\sum_{i=1}^{n_j+T_j}\tau_{jld_{ji}}h_{\t_j}(x_{ji},x_{j,i-1})\}.
\end{equation}

\medskip
\noindent {\bf 7.} 
The full conditional for $x_{j0}$ will be
\begin{equation}
\label{fullcondx0joint}
\Pi(x_{j0}\given\cdots)\propto\Pi(x_{j0})
\exp\left\{-\frac{\tau_{jld_1}}{2} h_{\t{j}}(x_{j1}, x_{j0})\right\}.
\end{equation}

\medskip
\noindent{\bf 8.} The full conditionals for the sampling of the out-of-sample observations 
for $k=1,\ldots,T_j-1$ are given by 
\begin{align}
\Pi(x_{j,n_{j}+k}| \cdots) & \propto \exp\left\{-\frac{1}{2}\left[\tau_{d_{j,n_j+k}}h_{\t_j}(x_{n_j+k},x_{n_j+k-1})\right.\right.\nonumber\\
\label{firstkminus1}
 & +\left.\left.\tau_{d_{j,n_j+k+1}}h_{\t_j}(x_{n_j+k+1},x_{n_j+k})\right]\right\}.
\end{align}
Also, for $k=T_j$, the full conditional is normal, with mean $g_j(\t_j,x_{j,n_j+T_j-1})$, and variance $\tau_{j\delta_{j,n_j+T_j}d_{j,n_j+T_j}}^{-1}$.

\medskip\noindent {\bf 9.} 
Having updated the selection probabilities to ${\bl p}^*$, and the geometric probabilities 
to ${\bl\l}^*$, we construct the geometric weights 
$(\pi_{jlk}^*)_{1\le j\le N^*}$ via equation (\ref{probweights1}).
Defining 
$$
\pi_{jl,N^*+1}^*=1-\sum_{k=1}^{N^*}\pi_{jlk}^*\quad{\rm and}\quad
\tau_{jl,N^*+1}^*\sim g_0,
$$
we are now ready to sample each $Z_{j,n_j+1}$ from its noise predictive 
\begin{align}
Z_{j,n_j+1} & \sim f_j(\,\cdot\,|{\bl x})=\sum_{l=1}^mp_{jl}^*
M_{jl}^*(\,\cdot\,|{\bl x}),\nonumber\\
 & {\rm with}\quad
   M_{jl}^*(z|{\bl x})
   =\sum_{k=1}^{N^*+1}\pi_{jlk}^*{\cal N}(z|0,(\tau_{jlk}^*)^{-1}).\nonumber
\end{align}
At each iteration of the Gibbs sampler, and for each $j$, we sample independently 
$\rho_j,\rho_j'\iid{\cal U}(0,1)$. So, $Z_{j,n_j+1}$, will be sampled from the a.s. 
finite mixture $M_{jl}^*(\,\cdot\,|{\bl x})$, with $1\le l\le m$, satisfying
$$
\sum_{r=0}^{l-1}p_{jr}^*<\rho_j\le\sum_{r=0}^lp_{jr}^*,\,\,\,p_{j0}^*=0,
$$
and from the $k$-th normal component ${\cal N}(\,\cdot\,|0,(\tau_{jlk}^*)^{-1})$ 
of the $M_{jl}^*(\,\cdot\,|{\bl x})$ mixture, with $1\le k\le N^*+1$, satisfying
$$
\sum_{r=0}^{k-1}\pi_{jlr}^*<\rho_j'\le\sum_{r=0}^k\pi_{jlr}^*,\,\,\,\pi_{jl0}^*=0.
$$

Details on sampling efficiently via embedded Gibbs samplers, thus circumventing Metropolis-within-Gibbs implementations for the nonstandard densities arising in equation (\ref{fullcondthetajoint}) through
(\ref{firstkminus1}), are provided in the supplementary file of \cite{merkatas2017bayesian}.

\providecommand{\noopsort}[1]{}\providecommand{\singleletter}[1]{#1}%

\end{document}